\documentclass[fleqn,usenatbib]{mnras}

\usepackage{newtxtext,newtxmath}
\usepackage[T1]{fontenc}
\usepackage{ae,aecompl}

\usepackage{graphicx}	% Including figure files
\usepackage{subcaption} % Enables subfigures
\usepackage{booktabs}% Advanced tables
\usepackage{multirow}
\usepackage{cleveref}

\title[Global 21cm Bayesian Foreground Modelling]{A General Bayesian Framework for Foreground Modelling and Chromaticity Correction for Global 21cm Experiments}

\author[D. Anstey et al.]{
Dominic Anstey,$^{1}$\thanks{E-mail: \href{da401@mrao.cam.ac.uk}{da401@mrao.cam.ac.uk}}
Eloy de Lera Acedo$^{1}$\thanks{E-mail: \href{eloy@mrao.cam.ac.uk}{eloy@mrao.cam.ac.uk}}
 and Will Handley$^{1,2}$\thanks{E-mail: \href{wh260@mrao.cam.ac.uk
}{wh260@mrao.cam.ac.uk}}
\\
$^{1}$Astrophysics Group, Cavendish Laboratory, J. J. Thomson Avenue, Cambridge, CB3 0HE, UK\\
$^{2}$Kavli Institute for Cosmology, Madingley Road, Cambridge, CB3 0HA, UK\\
}

\date{Accepted XXX. Received YYY; in original form ZZZ}

\pubyear{2020}

\begin{document}
\label{firstpage}
\pagerange{\pageref{firstpage}--\pageref{lastpage}}
\maketitle

\begin{abstract}
 The HI 21cm absorption line is masked by bright foregrounds and systematic distortions that arise due to the chromaticity of the antenna used to make the observation coupling to the spectral inhomogeneity of these foregrounds. We demonstrate that these distortions are sufficient to conceal the 21cm signal when the antenna is not perfectly achromatic and that simple corrections assuming a constant spatial distribution of foreground power are insufficient to overcome them. We then propose a new physics-motivated method of modelling the foregrounds of 21cm experiments in order to fit the chromatic distortions as part of the foregrounds. This is done by generating a simulated sky model across the observing band by dividing the sky into $N$ regions and scaling a base map assuming a distinct uniform spectral index in each region. The resulting sky map can then be convolved with a model of the antenna beam to give a model of foregrounds and chromaticity parameterised by the spectral indices of the $N$ regions.  We demonstrate that fitting this model for varying $N$ using a Bayesian nested sampling algorithm and comparing the results using the evidence allows the 21cm signal to be reliably detected in data of a relatively smooth conical log spiral antenna. We also test a much more chromatic conical sinuous antenna and find this model will not produce a reliable signal detection, but in a manner that is easily distinguishable from a true detection. 
\end{abstract}


%%%%%%%%%%%%%%%%%%%%%%%%%%%%%%%%%%%%%%%%%%%%%%%%%%

%%%%%%%%%%%%%%%%% BODY OF PAPER %%%%%%%%%%%%%%%%%%

\section{Introduction}\label{sec:intro}
The details of the development of the universe between the epoch of recombination and the formation of modern structure are not well understood. This is because this intermediate period of the cosmic dark ages, cosmic dawn and epoch of reionisation is very difficult to detect directly. External observations, such as the lack of hydrogen continuum absorption in high redshift galaxies \citep[e.g][]{schneider91,franx97} and the size of 10 degree scale anisotropies in the CMB \citep[e.g][]{knox98} have constrained the cosmic dawn and epoch of reionisation to a redshift range of $\sim 5 - 50$. However, no confirmed, direct detection has yet been made.

One of the most promising mechanisms by which this period may be probed is through use of the hyperfine absorption of neutral hydrogen at 21cm. Neutral hydrogen gas in the universe will absorb from, (or emit into), the radio background at 21cm in its rest frame, producing a change in temperature relative to the background that varies with redshift and observing direction. There are two main approaches currently being taken to attempt a detection of this hydrogen 21cm signal. The first is with interferometric instruments like the upcoming SKA \citep{dewdney09}, as well as HERA \citep{deboer17}, PAPER \citep{parsons10}, MWA \citep{lonsdale09} and LOFAR \citep{vanhaarlem13}. These experiments are designed to detect the full spatially varying power spectrum of the 21cm signal.

Alternatively, there are experiments designed to detect the spatially averaged ``monopole'' 21cm signal. These ``global'' 21cm experiments include EDGES \citep{bowman08}, SARAS \citep{patra13, singh18}, PRIZM \citep{philip19}, SCI-HI \citep{voytek14}, LEDA \citep{price18}, DAPPER (\href{https://www.colorado.edu/ness/dark-ages-polarimeter-pathfinder-dapper}{https://www.colorado.edu/ness/dark-ages-polarimeter-pathfinder-dapper}), MIST (\href{http://www.physics.mcgill.ca/mist/}{http://www.physics.mcgill.ca/mist/}) and BIGHORNS \citep{sokolowski15}. The techniques discussed in this paper are being designed for use in the experiment REACH\footnote{\url{https://www.astro.phy.cam.ac.uk/research/research-projects/reach}} \citep{acedo19}. However, they are applicable to any global 21cm experiment. 

The first detection of a global 21cm was reported recently by EDGES. \citet{bowman18} claimed a detection of an absorption trough centred at 78MHz. The identified signal is particularly notable for its depth, which is around 0.5K. This is much deeper than is permitted by current cosmological and astrophysical models \citep{cohen17,cohen20}. There are two main ways this can be explained. One is that the hydrogen gas cools more than in standard cosmological models due to interactions with dark matter \citep[e.g][]{munoz18,berlin18,barkana18a,barkana18b,slatyer18,liu19}. The other is that there is enhanced radio background beyond the CMB \citep[e.g.][]{bowman18,ewall18,ewall20,feng18,fialkov19,mirocha19}. Both of these cases enhance the difference between the radio background temperature and the hydrogen gas temperature, which enables a deeper signal, as will be explained in \Cref{sec:21cm_cosm}.

However, examination of the publicly available, processed EDGES data\footnote{\url{https://loco.lab.asu.edu/edges/edges-data-release/}}, such as that performed in \citet{hills18}, has indicated that there may be unaccounted for systematic errors in the data that could distort or mask the signal. Recently, \citet{sims19} showed that the presence of a damped sinusoidal systematic is strongly preferred in the EDGES data. \citet{bevins20} and \citet{singh19} have also demonstrated the possibility of residual systematics in the data. These systematics could have a wide range of sources, including residual beam effects, which are investigated further in this paper and in, for example, \citet{tauscher20a}, or ionospheric distortion, as discussed in \citet{shen21}.

\subsection{Background}\label{sec:background}
\subsubsection{Hydrogen 21cm Cosmology}\label{sec:21cm_cosm}
The change in temperature relative to the background radiation produced by hydrogen 21cm absorption varies with redshift, $z$, as
\begin{multline}
    \delta T(\nu) \approx 23(1 + \delta) x_\mathrm{H}(z) \left(\frac{T_\mathrm{S}-T_{\gamma}}{T_\mathrm{S}} \right) \left(\frac{\Omega_\mathrm{b}h^{2}}{0.02}\right) \\ \times \left[\left(\frac{0.15}{\Omega_\mathrm{m}h^{2}}\right)\left(\frac{1+z}{10}\right)\right]^{\frac{1}{2}} \mathrm{mK},
\end{multline}
where $T_\mathrm{S}$ and $T_\gamma$ are the hydrogen spin temperature and radio background brightness temperature respectively, $1+\delta$ is the local baryon overdensity, $x_\mathrm{H}(z)$ is the neutral fraction of hydrogen and $\Omega_\mathrm{m}h^{2}$ and $\Omega_\mathrm{b}h^{2}$ are the total matter and baryon densities respectively \citep{zaldarriaga04}.

\begin{figure}
    \centering
    \includegraphics[width=\columnwidth]{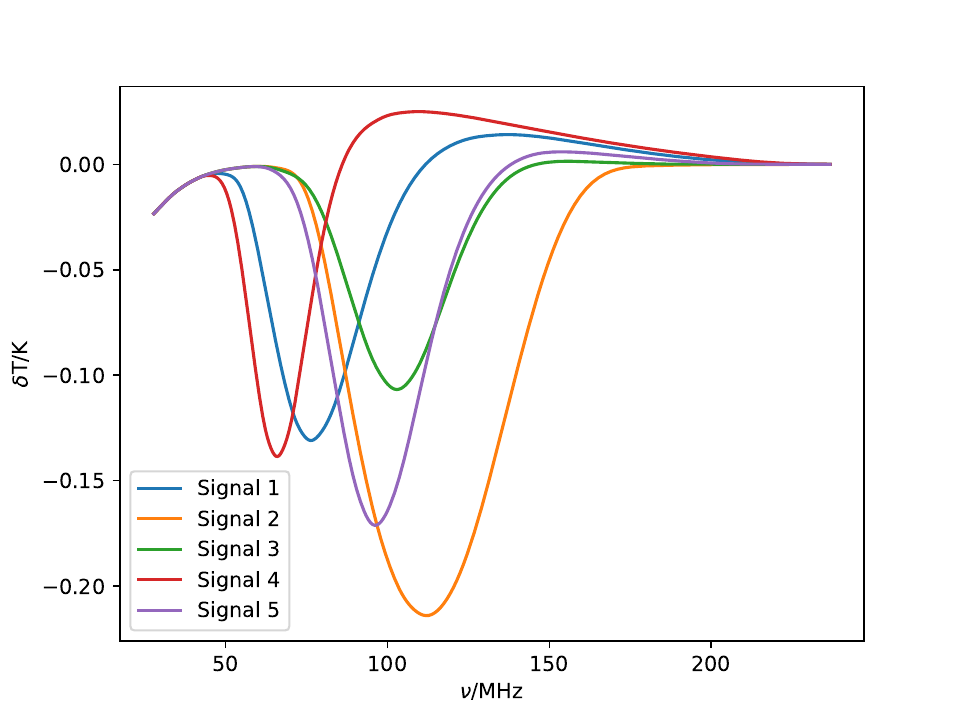}
    \caption{Five examples of possible global 21cm absorption troughs, generated using the ``21cmGEM' simulation tool \citep{cohen20}, using the astrophysical parameters given in \Cref{table:signal_params}.}
    \label{fig:example_21cm}
\end{figure}

The Wouthuysen-Field effect \citep{wouthuysen52,field58} will drive $T_\mathrm{S}$ towards $T_\mathrm{K}$, the kinetic temperature of the HI gas, in the presence of Lyman-alpha photons produced by stars during cosmic dawn. As $T_\mathrm{K}$ is below the background temperature $T_\gamma$, this results in an absorption relative to the background. X-ray heating then decreases the absorption intensity and may push the 21cm line into emission, before reionisation takes $x_\mathrm{HI}$ to zero and eliminates the signal \citep{furlanetto16}. This process should produce a detectable absorption trough that encodes information about the cosmic dawn and epoch of reionisation. Some examples of this absorption trough, for a range of different astrophysical parameters as specified in \Cref{table:signal_params}, are shown in \Cref{fig:example_21cm}. For the remainder of this work, we will approximate these absorption troughs using a Gaussian for simplicity. More detailed physical modelling of more realistic signals will be considered in future work.

\begin{table}
    \centering
    \begin{tabular}{|c|c|c|c|c|c|c|c|}
        \hline
         Signal & $f_*$ & $V_\mathrm{c}/\mathrm{kms}^{-1}$ & $f_\mathrm{x}$ & $\tau$ & $\alpha$ & $\nu_{\mathrm{min}}/\mathrm{keV}$ & $R_{\mathrm{mfp}}/\mathrm{Mpc}$\\
         \hline
         1 & 0.02 & 16.5 & 1 & 0.06 & 1 & 0.2 & 30 \\
         2 & 0.3 & 71.4 & 0.01 & 0.76 & 1.2 & 2.7 & 12 \\
         3 & 0.02 & 50.2 & 2.5 & 0.07 & 1.1 & 0.7 & 27 \\
         4 & 0.32 & 21.9 & 9.4 & 0.057 & 1.3 & 1.6 & 32 \\
         5 & 0.48 & 69.9 & 8.7 & 0.063 & 1.1 & 2.1 & 20 \\
         \hline
    \end{tabular}
    \caption{The astrophysical parameters used to generate the signals shown in \Cref{fig:example_21cm}. $f_*$ is defined as the star formation efficiency, $V_\mathrm{c}$ as the minimum circular velocity of star forming halos, $f_\mathrm{x}$ as the radiation efficiency of X-rays, $\tau$ as the optical depth of the CMB, $\alpha$ as the power law slope of X-ray radiation, $\nu_{\mathrm{min}}$ as the cut-off of the X-ray spectral energy distribution and $R_{\mathrm{mfp}}$ as the mean free path of ionising photons \citep{cohen20}.}
    \label{table:signal_params}
\end{table}

One of the primary difficulties in detecting this signal is the presence of extremely bright foregrounds in the relevant frequency band. The 21cm absorption trough is expected to have a depth of the order of $\sim 0.1\mathrm{K}$. However, the radio foregrounds present at these frequencies can exceed this by more than four orders of magnitude, making the signal very hard to distinguish. 

In much of the existing work on global 21cm experiments, the foregrounds are distinguished from the 21cm signal based on spectral differences. As the foregrounds are predominantly synchrotron and free-free radiation \citep{shaver99}, they will be spectrally smooth. Conversely, the 21cm signal is expected to be much less smooth. Therefore, the two components should be distinguishable by fitting a smooth function for the foregrounds, which will model the foregrounds but not the 21cm signal. Polynomial or log-polynomial functions can fulfil this purpose.

However, this is impeded by the effects of the chromaticity of the antenna used. Variations in the beam directivity pattern with frequency couple the spatial structure of power across the sky into the frequency domain. This results in non-smooth ``chromatic distortion'' arising in the foregrounds. These distortions, as will be demonstrated in \Cref{sec:polyfore_fit}, are degenerate with the 21cm signal, and so result in false or inaccurate signal detections, or prevent the signal from being identified at all \citep{tauscher20a}.

In this work, we will investigate the effects of systematic chromaticity distortions and discuss a method by which they can be modelled directly in a physically interpretable sense.

\subsubsection{Bayesian Inference}\label{sec:bayes}
Bayesian inference is a statistical model fitting technique, in which a model $\mathcal{M}$, parameterised by $\theta_\mathcal{M}$, is fit to a set of data, $\mathcal{D}$, by updating previous knowledge of the parameters, known as the prior, based on information from the data and any knowledge of the physical system. This is done by invoking Bayes' theorem, which states

\begin{equation}\label{eq:bayes_full}
        \mathrm{P}\left(\theta_\mathcal{M} | \mathcal{D}, \mathcal{M}\right) = \frac{\mathrm{P}\left(\mathcal{D} | \theta_\mathcal{M} \mathcal{M}\right)\mathrm{P}\left(\theta_\mathcal{M} | \mathcal{M}\right)}{\mathrm{P}\left(\mathcal{D}|\mathcal{M}\right)}.
\end{equation}
This can be written more succinctly as
\begin{equation}\label{eq:bayes_abbrv}
        \mathcal{P} = \frac{\mathcal{L}\mathcal{\pi}}{\mathcal{Z}},
\end{equation}
where $\mathcal{\pi}$ describes the prior probability distribution of the parameters and $\mathcal{P}$ describes the posterior distribution. $\mathcal{L}$ is the likelihood,which describes the probability of observing the data given a particular model and set of parameters for that model. $\mathcal{Z}$ is the Bayesian evidence. This can be defined as the probability of observing the data given a particular model, marginalised over all possible parameter values that model could take,

\begin{equation}\label{eq:evidence}
    \mathcal{Z} = \int \mathrm{P}\left(\mathcal{D} | \theta_\mathcal{M} , \mathcal{M}\right) \mathrm{P}\left(\theta_\mathcal{M} | \mathcal{M}\right) d\theta_\mathcal{M} = \int \mathcal{L}\mathcal{\pi} d\theta_\mathcal{M}.
\end{equation}

Bayesian inference is useful for two main tasks; estimating the parameter values for a given model that give the best agreement to the data (parameter estimation) and evaluating how well a given model can describe the data overall (model comparison). Parameter estimation follows \Cref{eq:bayes_full} to calculate the parameter's posterior distribution from the known prior distribution and likelihood. Many algorithms have been developed for this purpose, predominantly using Markov Chain Monte-Carlo (MCMC) methods, such as the relatively simple Metropolis-Hastings (e.g. \citet{chib95}) or the more sophisticated emcee v3 \citep{foremanmackey13}.

When performing parameter estimation directly, the evidence $\mathcal{Z}$ can simply be considered a normalisation constant for the posterior distribution and is not calculated in most MCMC methods. However, in order to perform model comparison, the evidence becomes critical. The probability of a particular model given the data can be derived from the evidence by applying Bayes' theorem again

\begin{equation}\label{eq:model_comp}
    \mathrm{P}\left(\mathcal{M}|\mathcal{D}\right) = \frac{\mathrm{P}\left(\mathcal{D}|\mathcal{M}\right)\mathrm{P}\left(\mathcal{M}\right)}{\mathrm{P}\left(\mathcal{D}\right)} = \mathcal{Z}\frac{\mathrm{P}\left(\mathcal{M}\right)}{\mathrm{P}\left(\mathcal{D}\right)}.
\end{equation}
$\mathrm{P}\left(\mathcal{D}\right)$ is a normalisation factor independent of the model. Therefore, the relative probabilities of two different models describing the data is simply given by the ratio of the two evidences, weighted by the model priors $\mathrm{P}\left( \mathcal{M}\right)$. Models will typically be assigned uniform weightings when compared in this way.

Calculating the evidence of an $n-$parameter model, as described in \Cref{eq:evidence}, requires integrating over an $n-$dimensional parameter space, where typically the majority of the posterior mass is contained within a small fraction of the prior volume. This quickly becomes prohibitively difficult as $n$ increases. Instead, the evidence can be estimated by the nested sampling algorithm, proposed in \citet{skilling06}.

Nested sampling is an algorithm designed with the primary goal of estimating the Bayesian evidence. However, the algorithm's process also produces samples from the posterior, which allows parameter estimation to also be achieved as a by-product. 

\citet{trotta08} gives an overview of the use of Bayesian data analysis in cosmology. However, the application of Bayesian analysis to the field of global 21cm cosmology has only begun relatively recently. It is now being used to analyse, for example, the data of LEDA \citep{bernardi16}, SARAS \citep{singh18} and EDGES high \citep{monsalve19}. Recently, \citet{sims19} also performed an extensive, fully Bayesian analysis on the publicly available, processed EDGES low data \citep{bowman18} to test for calibration systematics.

In this paper, we will propose a foreground modelling technique that makes heavy use of the Bayesian evidence to set its complexity. This means a nested sampling algorithm that gives an estimate of the evidence is necessary to properly implement this model. In our analysis, we use \texttt{PolyChord} \citep{handley15a,handley15b} for this purpose. 

\texttt{PolyChord} is an implementation of nested sampling that uses slice sampling to generate new live points from the previous ones. It has numerous tuneable settings that alter how the fit is performed. These are described in \cite{handley15a}. nlive is the number of samples maintained during the run. It controls how thoroughly the parameter space is sampled. num\_repeats controls the number of repeats of the slice sampling procedure, determining how correlated each new live point is to the previous ones. Increasing it improves the reliability of the calculated evidence. nprior and nfail are the number of samples initially drawn from the prior and the number of consecutive failed sample draws at which the algorithm should terminate, respectively. precision\_criterion also controls the algorithm termination. It is the fraction of the total evidence volume that the live points should occupy before termination. boost\_posterior controls how many additional posterior samples should be drawn and do\_clustering determines whether the algorithm should treat different modes partially independently.

\texttt{PolyChord} is preferable in this case to the primary alternative nested sampling algorithm \texttt{MULTINEST} \citep{feroz09} due to \texttt{PolyChord} performing more efficiently for the high model dimensionalities required by this problem. \texttt{PolyChord} also has the benefit of allowing ranking of parameter speeds to be exploited. While this functionality is not used in this work, the capacity to rank parameters by speed is beneficial for future extensions of the method, such as if a more detailed cosmological simulation is used for the 21cm signal modelling, rather than the simple Gaussian used here. 

This paper will be ordered as follows. In \Cref{sec:chrom_effects}, we analyse the effects of chromaticity of the antenna on global 21cm observations and the extent to which they will impede detections of the 21cm signal. In \Cref{sec:newmodel}, we present a new method of modelling foregrounds, using physical data of the sky and antenna beam to account for chromatic distortions. In \Cref{sec:results}, we show the results of testing this new foreground model on simulated data sets and explore its limits and in \Cref{sec:conclusions} we conclude the paper.

\section{Characterising the Effect of Chromaticity}\label{sec:chrom_effects}
In this section, we investigate the extent of the effects of antenna chromaticity on the ability to detect a global 21cm signal. In \Cref{sec:sky_sim}, we describe a simple but realistic sky simulation used to generate simulated foreground data affected by antenna chromaticity. In \Cref{sec:polyfore_fit}, we demonstrate the degree of distortions chromaticity can introduce into the foregrounds. In \Cref{sec:simple_correction}, we demonstrate the extent to which a simple chromaticity correction, such as that implemented in the EDGES result, cannot compensate for this effect.

\subsection{Sky Simulation}\label{sec:sky_sim}
In order to realistically model the chromatic distortion that can occur in a global 21cm experiment, the sky simulations were designed to have the following features,
\begin{enumerate}
    \item a realistic spatial power distribution
    \item a realistic pattern of spatial variation of spectral indices
\end{enumerate}
Ideally, the simulation would also model the second order effect  of spatial variation of spectral index curvature. However, as data at low frequencies is limited, it is very difficult to adequately model this. Therefore, investigation of this effect will be left for future work.

A map of spectral index variation across the sky was derived by calculating the spectral index required to map every pixel of an instance of the 2008 Global Sky Model (GSM) \citep{deoliveiracosta08} at 408MHz, $T_{408}\left(\theta,\phi\right)$, onto the corresponding pixel of an instance of the GSM at 230MHz, $T_{230}\left(\theta,\phi\right)$, according to
\begin{equation}\label{eq:B_map}
    \beta\left(\theta, \phi \right) = \frac{\log{\left(\frac{T_{230}\left(\theta,\phi\right)-T_\mathrm{CMB}}{T_{408}\left(\theta,\phi\right)-T_\mathrm{CMB}}\right)}}{\log{\left(\frac{230}{408}\right)}}.
\end{equation}

\begin{figure}
 \includegraphics[width=\columnwidth]{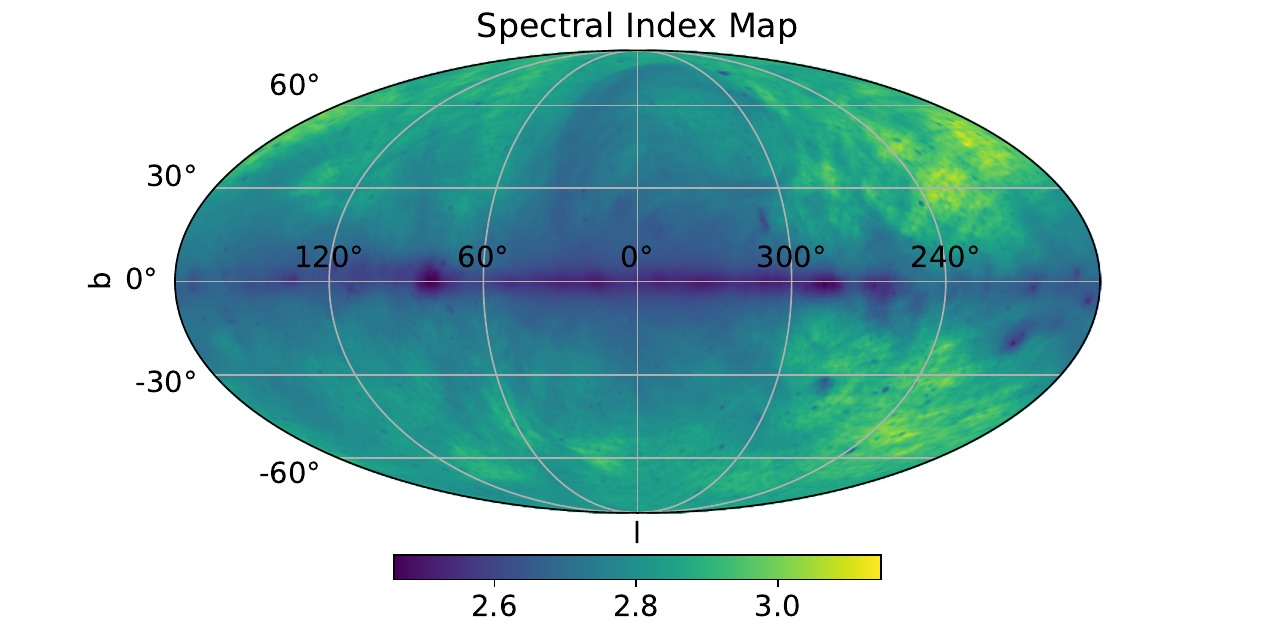}
 \caption{Map of spectral index variation across the sky in galactic co-ordinates. This map was generated by pixel-wise tracing between an instance of the 2008 Global Sky Model \citep{deoliveiracosta08} at 408MHz and an instance at 230MHz, according to \Cref{eq:B_map}.}
 \label{fig:B_map}
\end{figure}

A GSM instance at 230MHz was used to avoid using sky data in this calculation that may already be contaminated by the 21cm signal. \Cref{fig:B_map} shows the resulting spectral index map. 

This spectral index map was then used to generate a sky model according to 

\begin{equation}\label{eq:sky_model}
    T_\mathrm{sky}\left(\theta,\phi,\nu\right) = \left(T_\mathrm{408}\left(\theta,\phi\right) - T_\mathrm{CMB}\right)\left(\frac{\nu}{408}\right)^{-\beta\left(\theta,\phi\right)} + T_\mathrm{CMB}.
\end{equation}

For the purposes of this analysis, it is not required that this simulated sky be exactly accurate to the true sky. The simulation only needs to approximate the type of structures that would be seen on the true sky in order to determine their effects on the foreground data.

\begin{figure*}
    \centering
    \begin{subfigure}{\textwidth}
    \centering
    \includegraphics[width=0.3\textwidth]{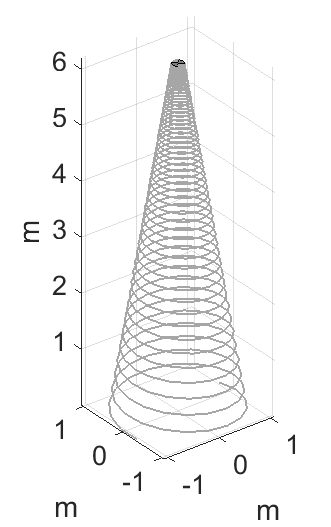}
    \centering
    \includegraphics[width=0.65\textwidth]{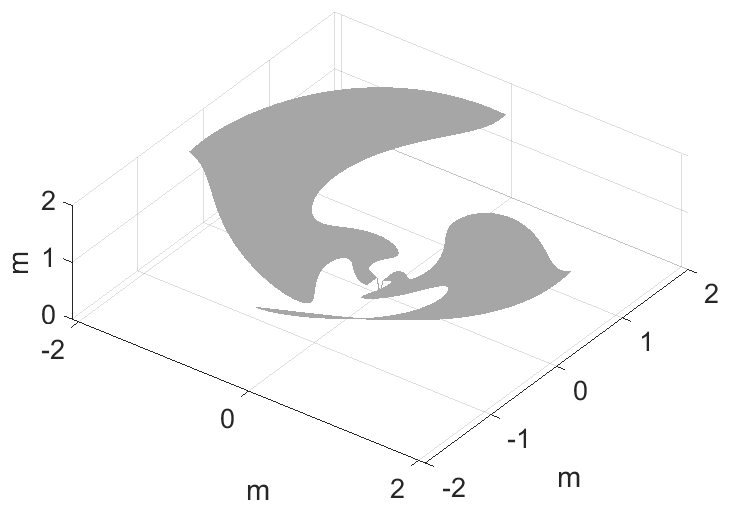}
    \end{subfigure}
    \caption{Diagrams of the log spiral antenna \citep{dyson65} (left), and the conical sinuous antenna \citep{buck08} (right), used in this analysis. Antenna patterns and plots provided by Quentin Gueuning and John Cumner.}
    \label{fig:antenna_diagrams}
\end{figure*}

\begin{figure*}
    \centering
    \includegraphics[width=\textwidth]{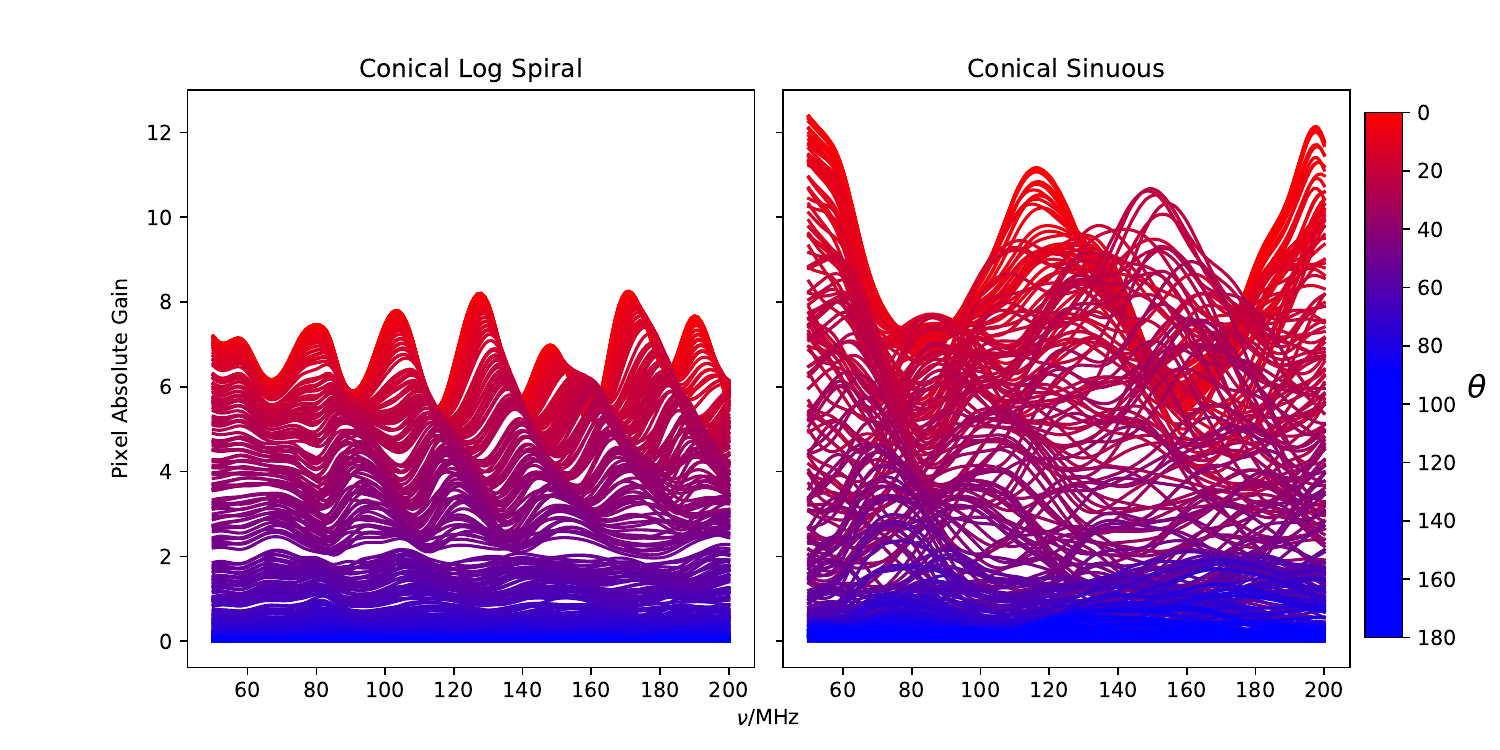}
    \caption{Measure of the beam patterns and chromaticity of the log spiral (left) and conical sinuous (right) antennae. Each line shows the variation in absolute directivity with frequency for a specific $\theta$ and $\phi$. Each plot shows the variation in directivity for 500 randomly chosen directions in the beam pattern, with the colour gradient showing the $\theta$ of that direction. The same 500 directions are used in each plot.}
    \label{fig:chromatic_plots}
\end{figure*}

Throughout this analysis, we will consider the chromatic distortions produced by a conical log spiral antenna \citep{dyson65}, of the form shown in \Cref{fig:antenna_diagrams} , which has a beam pattern as indicated in \Cref{fig:chromatic_plots}. This antenna was considered to simplify the results presented here. A simple dipole antenna may seem like a better choice. However, dipoles have a relatively narrow frequency band in which they are smooth, and then undergo beam splitting at higher frequencies. This variation in the degree of chromaticity across the band introduces an additional layer of complication to this problem, as the ability to fit a signal will vary with the centre frequency of the signal, relative to a log spiral antenna, which has an approximately uniform degree of chromaticity across the entire band. The details of this effect and the implications of the method presented here on choosing an antenna design for a global 21cm experiment will be considered in a later paper.

\subsection{Chromatic Residuals to Smooth Foreground Fits}\label{sec:polyfore_fit}
For a given Coordinated Universal Time (UTC), location and antenna pattern, a set of simulated data can be generated according to

\begin{equation}\label{eq:sim_data_gen}
    T_\mathrm{data}\left(\nu\right) = \frac{1}{4\pi}\int_{0}^{4\pi}D\left(\theta,\phi,\nu\right)\int_{t_\mathrm{start}}^{t_\mathrm{end}}T_\mathrm{sky}\left(\theta, \phi, \nu, t\right)dtd\Omega + \hat{\sigma}
\end{equation}
where $T_\mathrm{sky}\left(\theta, \phi, \nu, t\right)$ is as defined in \Cref{eq:sky_model}, rotated for time and observing location, and $D\left(\theta,\phi,\nu\right)$ describes the antenna's directivity pattern. It is assumed here that the antenna has a perfect ground plane such that no power is received by the antenna from the Earth. $\hat{\sigma}$ is uncorrelated Gaussian noise, which is used here for simplicity. 

In the case of a completely uniform spectral index, $\beta\left(\theta,\phi\right)=\beta$, and a perfectly achromatic antenna, $D\left(\theta,\phi,\nu\right) = D\left(\theta,\phi\right)$, \Cref{eq:sim_data_gen} reduces to

\begin{align}
    &T_\mathrm{data}\left(\nu\right) = A\left(\frac{\nu}{408}\right)^{-\beta} + T_\mathrm{CMB}, \\
    &A = \frac{1}{4\pi}\int_{0}^{4\pi}D\left(\theta,\phi\right)\int_{t_\mathrm{start}}^{t_\mathrm{end}}\left(T_\mathrm{408}\left(\theta,\phi,t\right)-T_\mathrm{CMB}\right)dtd\Omega.
\end{align}
Therefore, in this case, the data should be a perfect power law and fit well by a smooth function.

We characterise the chromatic distortion that arises due to beam chromaticity and non-uniform spectral index by fitting a smooth ``foreground'' function to data generated by \Cref{eq:sim_data_gen} and finding the residual. We use log polynomial functions of $5^\mathrm{th}$ order,

\begin{equation}\label{eq:log_poly}
    \log \left(T_\mathrm{model}(\nu)\right) = \sum_{i=0}^{4} a_{i}\log\left(\frac{\nu}{\nu_0}\right)^{i},
\end{equation}
with $\nu_0 = 140\mathrm{MHz}$.

This model is then fit to the data in a Bayesian sense with a likelihood function of 

\begin{equation}\label{eq:F_like}
    \log\mathcal{L} = \sum_{i}-\frac{1}{2}\log\left(2\pi\sigma_\mathrm{n}^{2}\right) - \frac{1}{2}\left(\frac{T_\mathrm{data}\left(\nu_{i}\right)-T_\mathrm{model}\left(\nu_{i}\right)}{\sigma_\mathrm{n}}\right)^{2}.
\end{equation}
This assumes a simple model of uniform uncorrelated Gaussian noise across the entire frequency band, which is the case for the simulated data being used here. The polynomial coefficient parameters $a_{i}$ were given very wide uniform priors of $\left[-10\mathrm{K}, 10\mathrm{K}\right]$ and the Gaussian noise parameter $\sigma_\mathrm{n}$ was given a logarithmically uniform prior of $\left[10^{-4}\mathrm{K}, 10^{1}\mathrm{K}\right]$.

The settings of \texttt{PolyChord} used in this fit, as defined in \Cref{sec:bayes}, are specified in \Cref{tab:polychord_settings}. These are the default settings and were used for all model fits in this paper unless otherwise stated.

\begin{table}
    \caption{\texttt{PolyChord} settings used in all model fits performed in this paper, as defined in \Cref{sec:bayes}. nDims is the dimensionality of the model. In general, for the models used here, $\mathrm{nDims} = n_\mathrm{foreground} + n_\mathrm{signal} + 1$. $n_\mathrm{foreground}$ depends on the model being used. $n_\mathrm{foreground} = 5$ for the polynomial foregrounds used in \Cref{sec:chrom_effects} and $n_\mathrm{foreground} = N$, where $N$ is the number of sky regions, for the physical model described in \Cref{sec:newmodel}. If a 21cm signal is included in the model, $n_\mathrm{signal} = 3$, using the 3-parameter Gaussian function shown in \Cref{eq:sig_model}. Otherwise, $n_\mathrm{signal} = 0$. The final parameter is the uncorrelated Gaussian noise, $\sigma_\mathrm{n}$.}
    \label{tab:polychord_settings}
    \centering
    \begin{tabular}{cc}
        \toprule
          nlive & $\mathrm{nDims}*25$ \\ \hline
          num\_repeats & $\mathrm{nDims}*5$ \\ \hline
          nprior & $\mathrm{nDims}*25$ \\ \hline
          nfail & $\mathrm{nDims}*25$ \\ \hline
          do\_clustering & True \\ \hline
          precision\_criterion & $0.001$ \\ \hline
          boost\_posterior & $0.0$ \\ \bottomrule
    \end{tabular}
\end{table}

\begin{figure*}
    \centering
    \includegraphics[width=\textwidth]{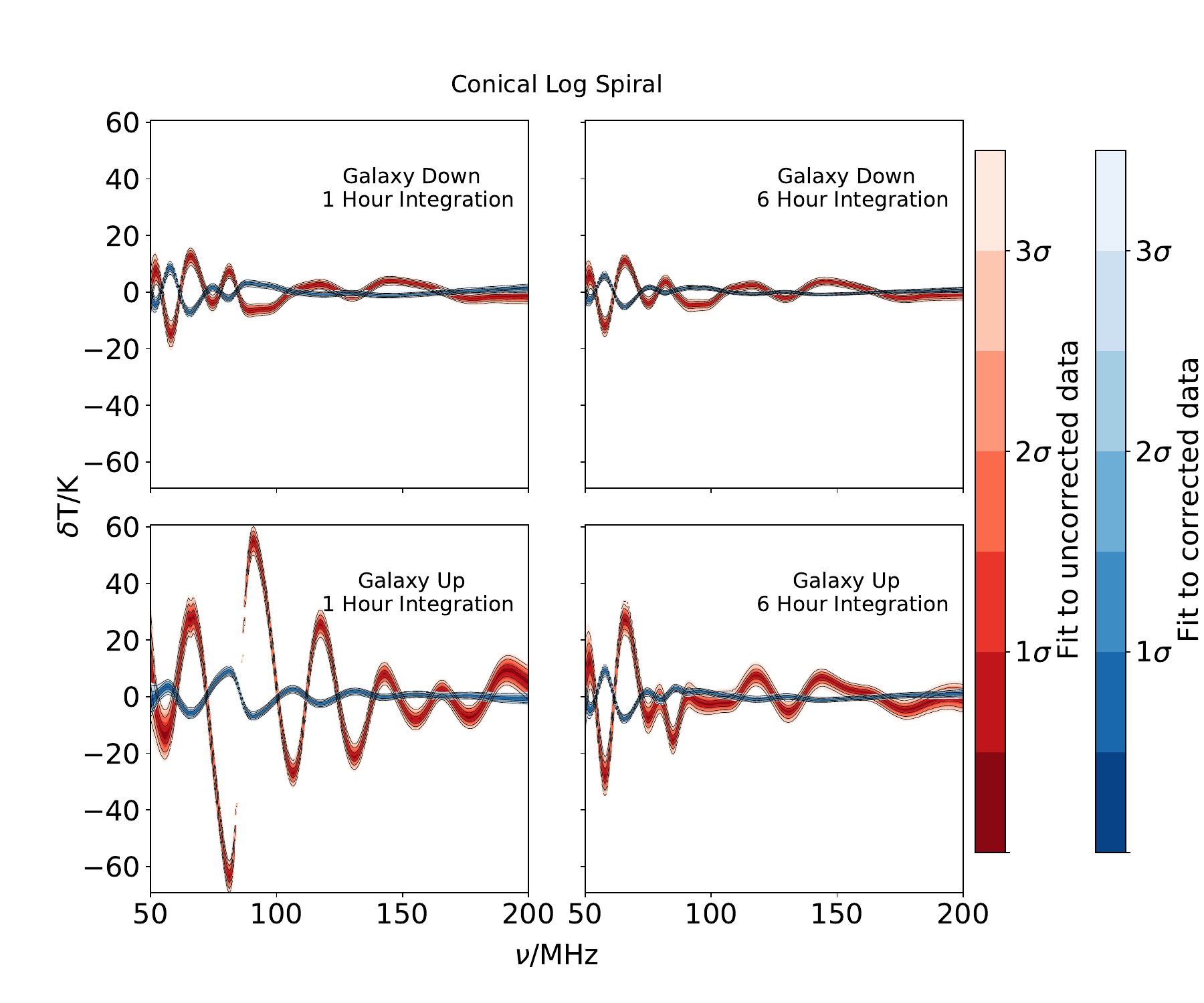}
    \caption{Residuals after subtraction of the best fit $5^\mathrm{th}$ order log polynomial from simulated sky data are shown in red. 'Galaxy Down' refers to simulated observations on 2019-10-01, when the galactic plane is below the horizon. 'Galaxy Up' refers to observations on 2019-07-01, with the galactic plane above the horizon. All observations begin at 00:00:00 and are integrated over the specified duration. Residuals after subtraction of the best fit $5^\mathrm{th}$ order log polynomial from simulated sky data that has been corrected for chromatic distortion as described in \Cref{eq:correction_div} are shown in blue. In all cases, the residuals greatly exceed the expected amplitude of the 21cm signal, even after the correction has been applied.}
    \label{fig:polyfit_residuals}
\end{figure*}

The chromatic distortion will prevent the detection of the 21cm signal if the distortions both have similar spectral structure to the 21cm signal and are similar or greater in magnitude.
\Cref{fig:polyfit_residuals} shows examples of the chromatic distortion for a logarithmic spiral antenna observing in the 50-200MHz band.

\begin{figure}
    \centering
    \includegraphics[width=\columnwidth]{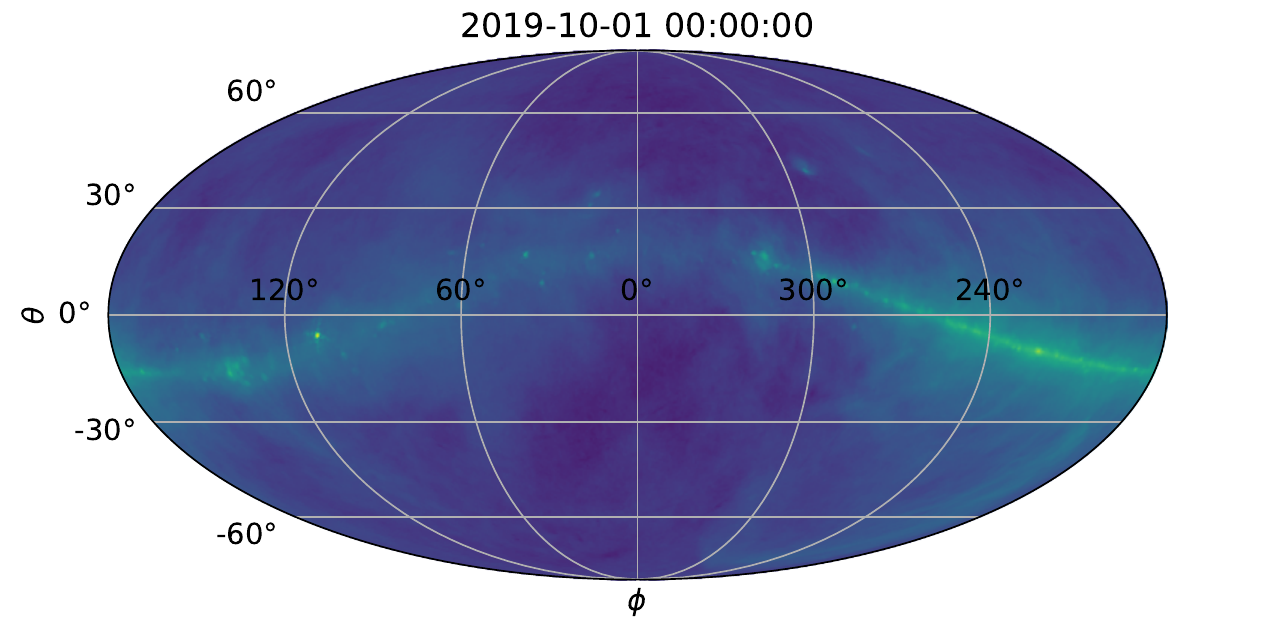}
    \hfill
    \includegraphics[width=\columnwidth]{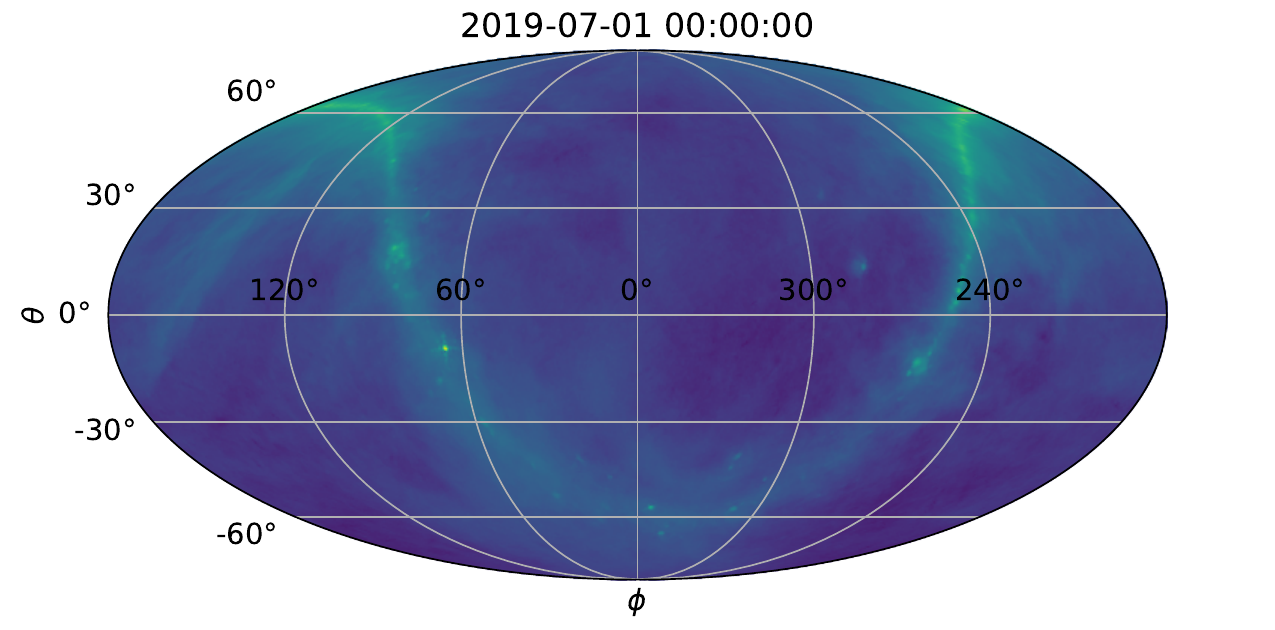}
    \caption{Plots of the overhead radio sky used to produce simulated observations when the galaxy is down (top) and when the galaxy is up (bottom). This assumes the antenna is located in the Karoo radio reserve in South Africa. This is shown in altitude ($\theta$) and azimuth ($\phi$) coordinates of the antenna's frame, with $\theta = 0$ marking the horizon.}
    \label{fig:overhead_skies}
\end{figure}

In order to demonstrate the effect of integration time and overhead structure on these chromatic distortions, four cases are considered. Two cases use simulated observations beginning at 2019-10-01 00:00:00 UTC, when the galactic disc is predominantly below or very close to the horizon for most of the night, for an antenna based at the Karoo radio reserve. For the rest of this analysis, these cases will be referred to as 'galaxy down' The other two use simulated observations beginning at 2019-07-01 00:00:00 UTC, when the galactic disc is predominantly above the horizon for most of the night. These will henceforth be referred to as 'galaxy up'. \Cref{fig:overhead_skies} shows the overhead structure in each case. For both observing times, 1 hour and 6 hours of integration are considered, with observations ending at 01:00:00 UTC and 06:00:00 UTC respectively. These correspond to local sidereal times (LSTs) of 2-3h and 2-8h for the cases where the galaxy is down, and 20-21h and 20-2h, for the cases where the galaxy is up.

If an observation is integrated over the same range of LST over multiple nights, the coupling of the chromatic antenna to the foreground power will be the same on each night. Therefore, the chromatic structure will be identical to that of a single night of observation. This assumes there is no change in the antenna's directivity over time. As we are only considering chromatic structure here, and other effects that would benefit from longer integration, such as noise reduction, RFI flagging and ionospheric distortion, are not included, only 1 night of observation is needed to reach the limit of the effect that time integration can have on this process, without requiring separate observations to be taken weeks or months apart. We therefore take 6 hours as the upper limit of observing time that could be reasonably achieved in a single night. The case of using data from multiple observation periods separated by a long enough period to give significantly different chromatic structures will be considered in future work.

The results in \Cref{fig:polyfit_residuals} demonstrate that the effects of antenna chromaticity result in significant chromatic structure in the foregrounds in all cases. These distortions are, by definition, sufficiently non-smooth as to not be fitted by the smooth log-polynomial function and all have amplitudes much greater than the expected amplitude of the 21cm absorption trough, of $~10^{-1}\mathrm{K}$. As such, these chromatic distortions are more than sufficient to mask the 21cm signal if not properly corrected for. The observations when the galaxy is down do show much smaller chromatic distortion than when the galaxy is up, as should be expected. Furthermore, a longer integration time is seen to reduce the chromatic residuals to a degree in both cases. However, neither of these effects are sufficient to bring the amplitude of residuals to below the level of the 21cm signal. Therefore, these distortions are enough to mask the signal.

\subsection{Chromatic Residuals After a Uniform $\beta$ Correction}\label{sec:simple_correction}
This chromatic distortion could in theory be corrected for if both the antenna pattern and the foreground power distribution were exactly known. However, these are very difficult to know in practice, especially in the case of the foreground power. One possibility for a simple correction would be to assume an unchanging power distribution with frequency, or equivalently a uniform spectral index, across the observation band based on a known all-sky map, and correct for the distortions that such a model would imply. 

An example of this type of chromaticity correction is used for the EDGES data in, for example, \citet{mozdzen17} and \citet{mozdzen19}. In these analyses, a chromatic correction factor is calculated for each LST of the data from a base all-sky map and a simulated antenna beam that is assumed to be known exactly. The correction is of the form

\begin{equation}\label{eq:simple_correct}
    B_\mathrm{factor}\left(\nu, t\right) = \frac{\int_0^{4\pi}T_\mathrm{sky-model}\left(\theta,\phi,\nu_\mathrm{reference}, t\right)D\left(\theta, \phi, \nu\right)d\Omega}{\int_0^{4\pi}T_\mathrm{sky-model}\left(\theta,\phi,\nu_\mathrm{reference}, t\right)D\left(\theta, \phi, \nu_\mathrm{reference}\right)d\Omega},
\end{equation}
where

\begin{multline}\label{eq:fix_B_sky}
    T_\mathrm{sky-model}\left(\theta,\phi,\nu_\mathrm{reference}, t\right) = \\ \left[T_\mathrm{base}\left(\theta,\phi, t\right)-T_\mathrm{CMB}\right]\left(\frac{\nu_\mathrm{reference}}{\nu_\mathrm{base}}\right)^{-2.5}.
\end{multline}
Here, $T_\mathrm{base}\left(\theta,\phi, t\right)$ is a base all-sky map at a frequency $\nu_\mathrm{base}$, such as the Haslam map at 408MHz \citep{remazeilles15}, rotated to the appropriate time, $t$, and $\nu_\mathrm{reference}$ is a reference frequency within the observed band. 

The observed data at each LST can then be divided by its corresponding $B_\mathrm{factor}\left(\nu\right)$ to attempt to correct for chromatic distortions as in \Cref{eq:correction_div}. This correction makes the assumption that the power distribution of the sky across the entire band is well approximated by that of the base map, and that the sky power scales with a uniform spectral index of 2.5.

\begin{equation}\label{eq:correction_div}
    T_\mathrm{corr}\left(\nu\right) = \int_{t_\mathrm{start}}^{t_\mathrm{end}}\frac{T_\mathrm{data}\left(\nu, t\right) - T_\mathrm{CMB}}{B_\mathrm{factor}\left(\nu, t\right)}dt + T_\mathrm{CMB}.
\end{equation}

We test this simple correction model on our simulated data for a log spiral antenna. The correction was performed in five minute intervals for the duration of the 1 or 6 hour observing times.  

We used $T_{230}\left(\theta, \phi\right)$, which was defined in \Cref{eq:B_map} as $T_\mathrm{base}\left(\theta,\phi\right)$, for a base map that is near to but not in the observed band. As it was used in the generation of the simulated data, $T_{230}\left(\theta, \phi\right)$ is an accurate sky map at 230MHz for this simulation. $\nu_\mathrm{reference}$ was taken to be in the centre of the observing band, at 125MHz. 

The residuals after performing this correction on the simulated data sets described in \Cref{sec:sky_sim} and \Cref{sec:polyfore_fit} and fitting a $5^\mathrm{th}$ order log polynomial to the results are also shown in \Cref{fig:polyfit_residuals}. In comparison to the uncorrected residuals, performing the correction can be seen to reduce the magnitude of the residuals, especially in the case where the galactic plane was overhead. Furthermore, the evidences recorded in \Cref{tab:Z_comparison} show the corrected models to be strongly favoured over the uncorrected models in all cases. However, for the log spiral antenna tested here, this correction is still insufficient to reduce the residuals to a level at which a 21cm signal of amplitude $\sim 0.1-0.2K$ could be detected. It can also be seen that increasing the integration time gives less reduction in the residuals of this corrected data, when compared to the uncorrected case.

If the simulated data sets were generated by scaling $T_{230}\left(\theta, \phi\right)$ by a uniform 2.5 spectral index, as in \Cref{eq:fix_B_sky}, this correction would exactly remove the chromatic distortions in this test case. Therefore, the failure of this simple model to remove the distortions to a significant degree must arise from the varying spectral index used, which breaks the assumptions used in the simple correction. \cite{sims19} performed a thorough analysis of the uncorrected-for chromatic distortions, in the case where the beam pattern is not accurately known. However, an unknown foreground will compound these effects in a non-trivial manner.

\section{Foreground Modelling to Include Chromaticity}\label{sec:newmodel}
The results of \Cref{sec:chrom_effects} demonstrate that chromatic distortions will mask the 21cm signal in a global experiment. This is observed even in the ideal conditions of a reasonably smooth antenna when the galaxy is below the horizon. Furthermore, a simple chromaticity correction that does not account for the coupling of spectral index variation to the antenna chromaticity is found to be insufficient to fully correct for this distortion. 

This demonstrates that, due to variation in $\beta$, the effect of chromatic distortion is highly dependent on the precise form of the foregrounds observed by the antenna. Therefore, in order to properly correct for the distortion, a foreground model in which the chromatic distortions are derived alongside the foreground parameters is needed. 

We propose a model here which includes these distortions as part of the foreground by constructing a parameterised foreground out of parameterised sky maps and antenna patterns. An approximate all-sky map is generated for each observing frequency from a set of parameters. Similarly, an antenna directivity pattern is generated from parameters, and the two are convolved according to \Cref{eq:sim_data_gen} to produce a parameterised model that describes both the foregrounds and the chromatic distortion of the antenna. This model can then be fit to the data in a Bayesian sense, deriving the physically motivated parameters of the sky and antenna directly from the data and fitting the chromatic distortion as part of the foreground model.

In the following analysis, we assume the antenna pattern is known exactly and investigate the effects of physical modelling of the foregrounds alone. An analysis of the effects of an analogous parameterised antenna pattern will follow in a subsequent paper.

\subsection{The Spectral Index Model}\label{sec:B_model}
The parameterised sky model used in this process requires both a sufficiently accurate spatial power distribution and sufficiently accurate model of the spectral index variation, if it is to accurately model chromatic distortions. 

The known spatial power distribution is only required at a single frequency in or near the observed band, as if a sufficiently accurate model of spatially dependent spectral index is used, the power distributions at all other observed frequencies can be derived. For the following analysis, we have used the 230MHz instance of the 2008 GSM. However, using the 2015 reprocessed Haslam map \citep{remazeilles15} at 408MHz as the base map instead was found to produce very little difference in the results.

\begin{figure}
    \centering
        \includegraphics[width=\columnwidth]{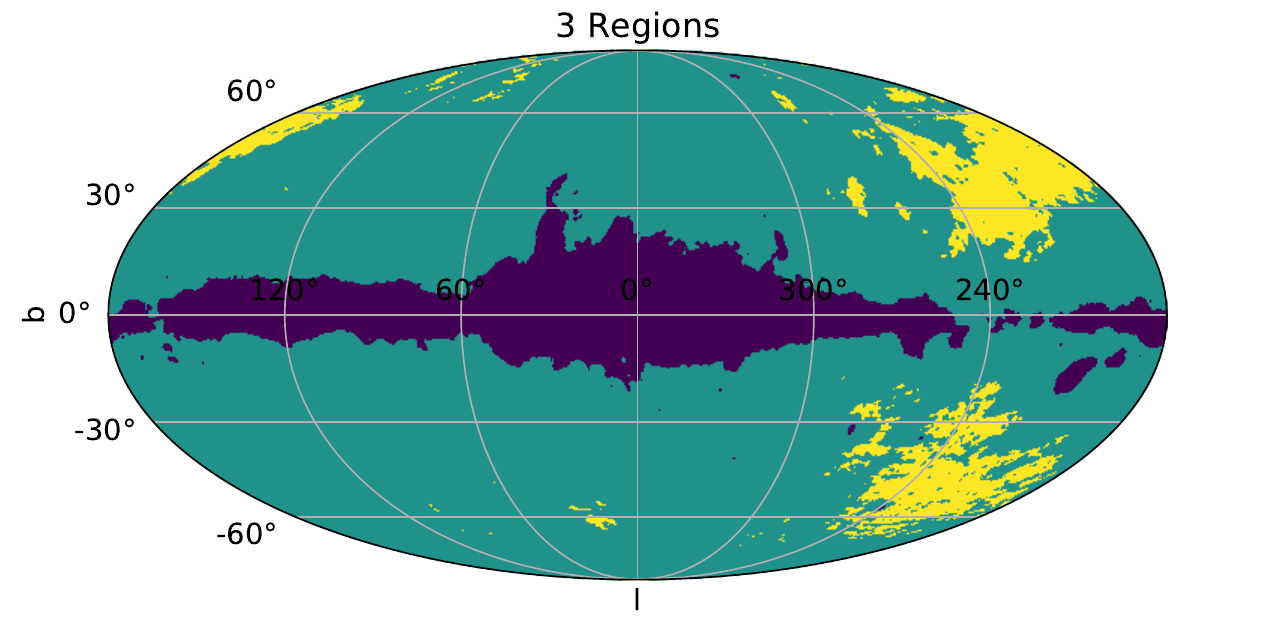}
    \hfill
        \includegraphics[width=\columnwidth]{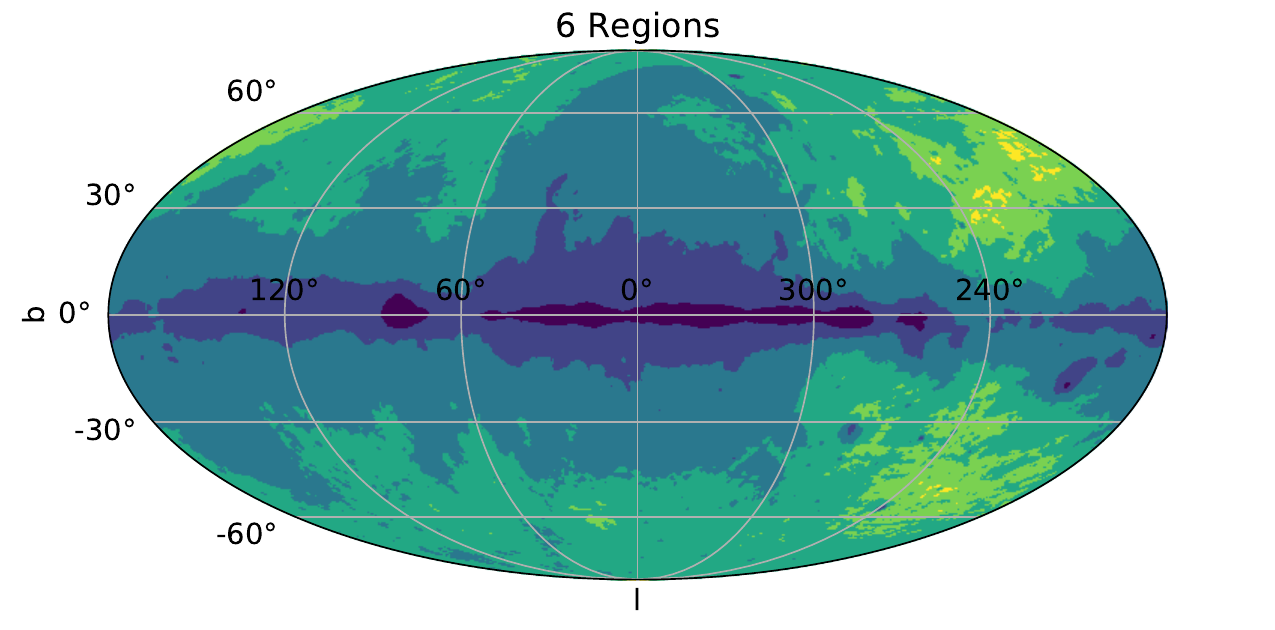}
    \caption{Division of sky into 3 regions (top) and 6 regions (bottom), in which the spectral indices of \Cref{fig:B_map} are similar, in galactic coordinates. For the 3 region division, the regions are the portions of the map shown in \Cref{fig:B_map} where the spectral indices are in the ranges 2.458-2.687, 2.687-2.917 and 2.917-3.146 respectively. Likewise, for the 6 region map, they are where the spectral indices are in the ranges 2.458-2.573, 2.573-2.687, 2.687-2.802, 2.802-2.917, 2.917-3.031 and 3.031-3.146. The physically motivated foreground model describes the sky by assuming a constant spectral index in each region.}
    \label{fig:region_maps}
\end{figure}

In order to model the spectral index distribution in a manner sufficiently approximate as to be parameterisable, while still maintaining sufficient accuracy, we used a process of coarse-grain dividing the spectral index map into $N$ regions, within which $\beta$ is similar. These regions were derived from the spectral index map shown in \Cref{fig:B_map} by dividing the total range of spectral indices into $N$ equal width sections, then defining each sky region as the area of the sky with spectral indices within each section. \Cref{fig:region_maps} shows the sky divisions that result when split into 3 and 6 regions.

An all-sky model at every observing frequency can then be calculated by assuming a distinct but uniform $\beta$ within each region and scaling the base map by power laws according to the resulting approximate $\beta$ map.

This model enables control over the complexity of the sky model and thus its potential to be accurate to the true sky. The assumption that $\beta$ is uniform within each region becomes more accurate as $N$ increases, since each region covers a smaller area on the sky, with less potential for $\beta$ to vary. Taking $N=1$ would be equivalent to assuming a constant uniform $\beta$ in the model. As $ N\xrightarrow{}\infty$, the model effectively assigns a unique spectral index to every pixel. However, an increase in $N$ requires a proportional increase in the number of parameters used. 

\subsection{Conversion to a Parameterised Foreground Model Function}\label{sec:fore_function}

With the sky divided into regions in this fashion, an all-sky model at each observing frequency can then be derived by propagating the base map in each region by a fixed spectral index. This results in a sky temperature of
\begin{multline}\label{eq:skymodel_update}
    T_\mathrm{sky}\left(\theta,\phi,\nu\right) = \\ \left[\sum_{i=1}^{N}M_i\left(\theta,\phi\right)\left(T_\mathrm{base}\left(\theta,\phi\right) - T_\mathrm{CMB}\right)\left(\frac{\nu}{230}\right)^{-\beta_i}\right] + T_\mathrm{CMB}.
\end{multline}
$M_i\left(\theta,\phi\right)$ describes a mask for the sky region, $i$, that has a value of 1 at every pixel within the defined region and 0 elsewhere. $\beta_i$ is the spectral index parameter of that sky region.

Integrating the sky temperature described in \Cref{eq:skymodel_update} with an antenna beam pattern as in \Cref{eq:sim_data_gen} can then produce a foreground model, which we define as

\begin{multline}\label{eq:unfac_foremodel}
    T_\mathrm{model}\left(\nu\right)  = \frac{1}{4\pi}\int_{0}^{4\pi}D\left(\theta,\phi,\nu\right)
    \\\times\int_{t_\mathrm{start}}^{t_\mathrm{end}}\sum_{i=1}^{N}M_i\left(\theta,\phi\right) \left(T_{230}\left(\theta,\phi\right) - T_\mathrm{CMB}\right)\left(\frac{\nu}{230}\right)^{-\beta_i}dtd\Omega 
    \\ + T_\mathrm{CMB}.
\end{multline}

As the integrals over $t$ and $\Omega$ are prohibitively computationally intensive to perform within a likelihood, we refactor this equation as

\begin{equation}\label{eq:fullforemodel}
    T_\mathrm{model}\left(\nu\right) = \left[\sum_{i=1}^{N}K_i\left(\nu\right)\left(\frac{\nu}{230}\right)^{-\beta_i}\right] + T_\mathrm{CMB},
\end{equation}
where 
\begin{multline}\label{eq:chromfunc}
    K_i\left(\nu\right) = \frac{1}{4\pi}\int_{0}^{4\pi}D\left(\theta,\phi,\nu\right)M_i\left(\theta,\phi\right) \\\times\int_{t_\mathrm{start}}^{t_\mathrm{end}}\left(T_{230}\left(\theta,\phi\right) - T_\mathrm{CMB}\right)dtd\Omega.
\end{multline}

This allows the $N$ ``chromaticity functions'', $K_i\left(\nu\right)$, to be precalculated outside of the likelihood, which greatly expedites likelihood evaluations.

\begin{figure*}
    \centering
        \includegraphics[width=\textwidth]{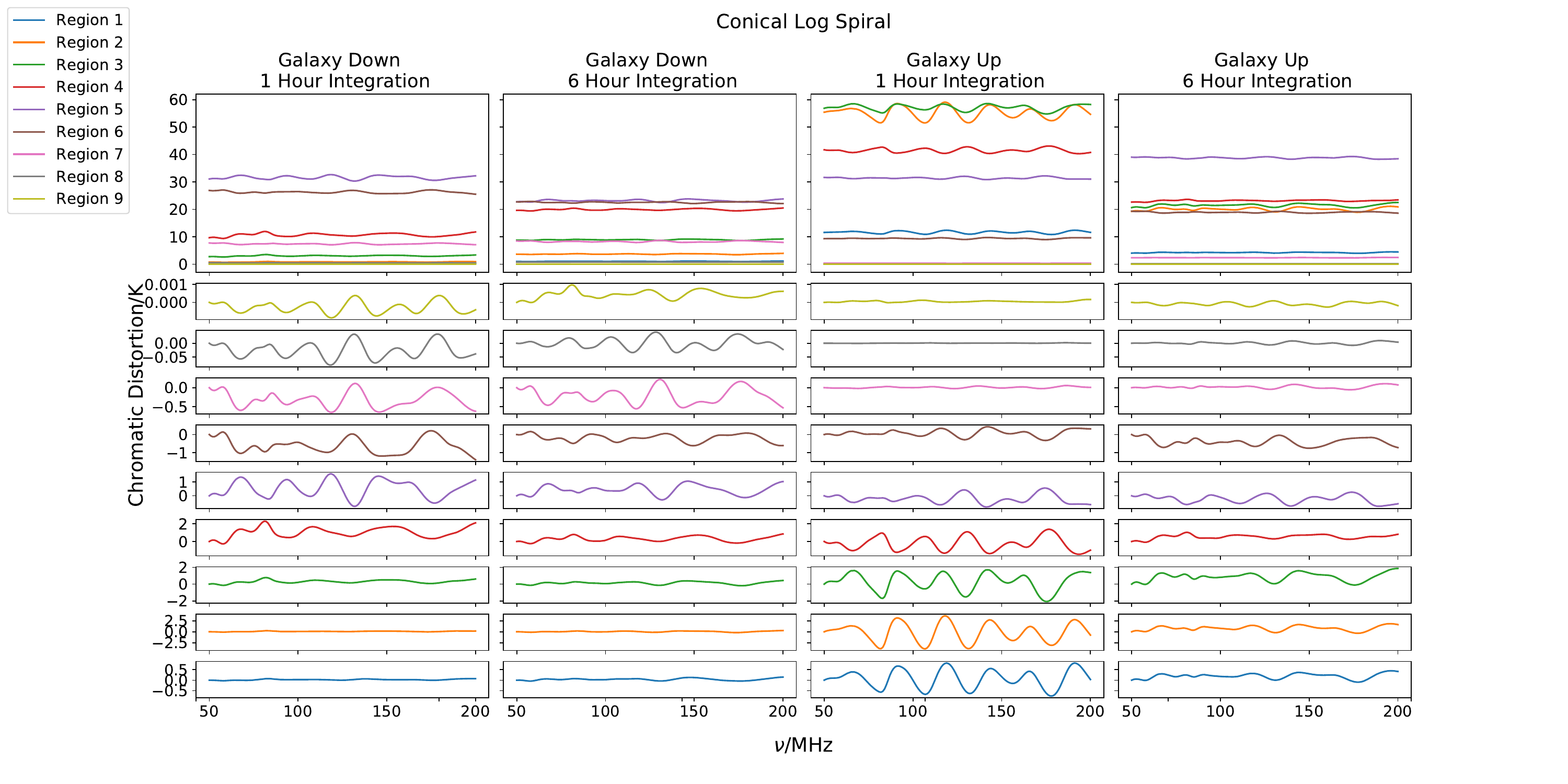}
    \caption{Chromatic functions of a log spiral antenna when $N=9$, as defined in \Cref{eq:chromfunc}. The top plots show the true absolute chromaticity functions. The bottom plots show the same functions shifted to be centred on zero and rescaled to show their variation.}
    \label{fig:log_spir_chromatic_functions}
\end{figure*}

\Cref{fig:log_spir_chromatic_functions} shows the chromaticity functions produced by a log spiral antenna, for 1 hour and 6 hours of observation, with the galaxy above and below the horizon. These plots are for $N=9$ in all cases, thus producing 9 chromaticity functions. The relative magnitude of each chromatic function is dependent on both the total power within the defined region and its overlap with the beam. $K_8\left(\nu\right)$ and $K_9\left(\nu\right)$ show much lower magnitudes in all cases due to covering much smaller areas. When the galaxy is below the horizon, the magnitude of $K_1\left(\nu\right)$ decreases, as could be expected. The chromaticity functions when the galaxy is up show a much greater degree of fluctuations than those when the galaxy is down. In addition, increasing the integration time results in the chromaticity functions smoothing somewhat and the difference between their absolute magnitudes decreasing. This is consistent with the results observed in \Cref{sec:polyfore_fit}, with less galactic contribution and longer integration time giving less chromatic distortion.

It should be noted that, in practice, there may be differences between the base map from which the model is generated, $T_{230}\left(\theta,\phi\right)$ in \Cref{eq:chromfunc}, and the true sky at that reference frequency, as the base map cannot be expected to be perfectly accurate. Therefore, one of the ways this model could be improved further would be to have it account for these potential differences. For example, each $K_i\left(\nu\right)$ could be multiplied by an additional scale factor parameter, $A_i$, with priors set by the uncertainties of the base map. This could then potentially be improved further by including correlations between these $A_i$ parameters. This extension to the model is not included here for simplicity, as the data in the test case considered here is generated from the same base map as is used in the data, so any $A_i$ parameters would be 1 by definition. However, when applying this method to real data, the analysis can be repeated using this addition, and similar models accounting for the uncertainty in the base map. The Bayesian evidence can then be used to determine which, if any, characterisations of base map uncertainty are favoured by the data.

This method could also potentially be enhanced further by considering multiple data sets taken at different observing times simultaneously. This would entail calculating a separate set of $K_i\left(\nu\right)$ chromatic functions for each relevant observation time period, resulting in separate $T_\mathrm{model}\left(\nu\right)$s for each. These can then be fit simultaneously to their corresponding data by using a log likelihood that is a sum of the log likelihood given in \Cref{eq:F_S_like}, using the same foreground and 21cm signal parameters in each model. Doing so would, in theory, provide additional information by which the foreground and signal parameters could be constrained more reliably. This effect of reduction in posterior error as more data times are considered is demonstrated in \citet{tauscher20a} and \citet{tauscher20b}. However, in this paper, we only consider the simpler case of using a single time-integrated data set and model, to demonstrate the technique.  
 
 Overall, this results in an $N$ parameter foreground model that contains information about both the variation in spectral index and the distortions due to chromaticity.
 
 \section{Results}\label{sec:results}
 In this section, we will discuss the outcomes of applying the proposed foreground modelling process to simulated data sets. In \Cref{sec:logspiral_results}, we will demonstrate the results of applying this model in the case of a log spiral antenna. In \Cref{sec:model_complexity}, we will discuss the use of the Bayesian evidence to select the model complexity. In \Cref{sec:limits}, will will discuss the limitations of this model when applied to a more complex antenna.
 
 \subsection{Application of Model to a Log Spiral Antenna}\label{sec:logspiral_results}
 This model was fit to the simulated data described in \Cref{sec:chrom_effects} for a log spiral antenna. A simulated Gaussian global 21cm signal was added to this data, 
 
 \begin{equation}\label{eq:sig_model}
     T_{21}\left(\nu\right) = -A\exp{\left(-\frac{\left(\nu-\nu_\mathrm{c}\right)^{2}}{2\sigma^2}\right)},
 \end{equation}
 with amplitude $A=0.155\mathrm{K}$, centre frequency $\nu_\mathrm{c} = 85\mathrm{MHz}$ and width $\sigma = 15\mathrm{MHz}$.

The physically motivated foreground model was fit to this data jointly with a Gaussian signal model, using the likelihood

\begin{multline}\label{eq:F_S_like}
    \log\mathcal{L} = \sum_{i}-\frac{1}{2}\log\left(2\pi\sigma_\mathrm{n}^{2}\right) \\ - \frac{1}{2}\left(\frac{T_\mathrm{data}\left(\nu_{i}\right)-\left(T_\mathrm{model}\left(\nu_{i}\right) + T_{21}\left(\nu_{i}\right)\right)}{\sigma_\mathrm{n}}\right)^{2}.
\end{multline}

In practice, the 21cm signal is more complicated than the Gaussian used here and would require a more detailed physically motivated model. However, we use a Gaussian here for simplicity to demonstrate the foreground modelling process.

The values of $\beta_{i}$ all have uniform priors over the range $[2.45844, 3.14556]$, which is the full range of spectral indices in the map in \Cref{eq:B_map}. The parameters of the signal model $T_{21}\left(\nu\right)$ had uniform priors of $[50\mathrm{MHz}, 200\mathrm{MHz}]$ for $\nu_c$, $[10\mathrm{MHz}, 20\mathrm{MHz}]$ for $\sigma$ and $[0\mathrm{K}, 0.25\mathrm{K}]$ for A. $\sigma_\mathrm{n}$ had a logarithmically uniform prior of $[10^{-4}\mathrm{K}, 10^{-1}\mathrm{K}]$.

This parameter space is quite complicated and highly multimodal, especially at higher $N$. Therefore, care must be taken when performing this fit in order to ensure the results are accurate. This was achieved by first running the fit with a high number of live points, in order to ensure that the optimal mode was found and thus the posteriors are accurate. The settings of \texttt{PolyChord} used in these initial runs are given in \Cref{tab:polychord_settings}, except with nlive = $\mathrm{nDims}*500$. Once this is done, a second 'enhancement' fit is performed on the same data, identical to the first in every respect except that it uses nlive = $\mathrm{nDims}*50$ and takes the priors of $\beta_{i}$ as uniform in the range $[\mu_i - 5\sigma_i, \mu_i + 5\sigma_i]$, where $\mu_i$ and $\sigma_i$ are the weighted means and standard deviations of the posteriors of the $\beta_i$ parameters of the initial run. If $\mu_i$ and $\sigma_i$ are such that this range exceeds that of the initial run, the prior limits of the initial run are used instead. This refocusing on the highest likelihood posterior peak helps ensure that the evidences are accurate, once the difference in prior is corrected for via a volume factor, by recalculating the evidence without having to search a highly multimodal parameter space. The details of these corrections will be discussed in a future work. 

\begin{figure*}
    \centering
    \includegraphics[width=\textwidth]{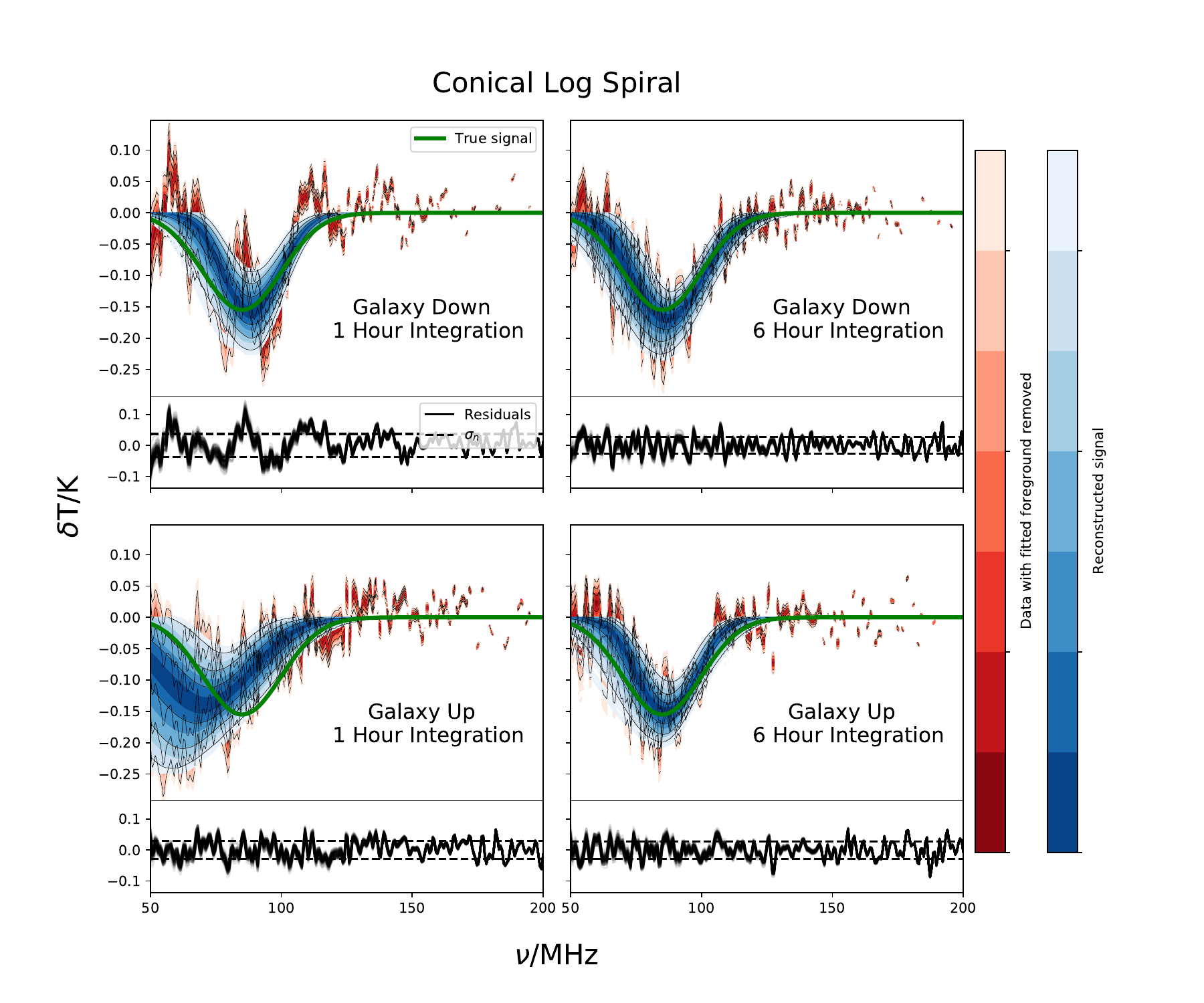}
    \caption{Results from fitting simulated observations with a log spiral antenna using the more detailed foreground model with $N=9$. Each plot fits the same simulated data set as the corresponding plot in \Cref{fig:polyfit_residuals}. The red region shows the residual after subtraction of the foreground model from the data, equivalent to the red region of \Cref{fig:polyfit_residuals}. The blue region shows the fitted 21cm signal and the green line shows the true 21cm signal that was added to the simulated data. Close agreement between all 3 curves indicates a well modelled foreground with few systematic residuals remaining and an accurately identified 21cm signal.}
    \label{fig:new_model_results}
\end{figure*}

\Cref{fig:new_model_results} shows the results of performing this fit on the same data sets described in \Cref{sec:chrom_effects}, for the log spiral antenna, with $N=9$.

The residuals after foreground subtraction are reduced by ~2-3 orders of magnitude compared to those using a log polynomial foreground shown in \Cref{fig:polyfit_residuals}. The chromatic distortions are thus fit sufficiently accurately by the improved model to enable a 21cm signal to be correctly identified. This is demonstrated by the strong agreement between the best fit 21cm model (blue) with the true signal (green). For both the case with the galaxy above the horizon and the case with it below, a longer integration time enables the signal to be modelled more accurately, as could be expected. However, at low integration times, the case with the galaxy above the horizon produces a more accurate fit, with much lower residuals, despite the fact that chromatic distortion is larger in this case compared to when the galaxy is down. This is likely due to the fact that the slightly larger chromatic structures here are less degenerate with the signal, enabling them to be more easily distinguished. However, for longer integration times, the two cases become almost entirely equivalent, showing a robustness of this technique to the time at which the observation takes place.

The improved quality of the model fitting compared to the previous methods discussed can be quantified using the Bayesian evidence. \Cref{tab:Z_comparison} details the evidences of the fits of all 4 data sets with a $5^\mathrm{th}$ order log polynomial, $5^\mathrm{th}$ order log polynomial after a uniform $\beta$ chromatic correction and with the physically motivated model for $N=9$.

\begin{table*}
    \caption{Table of evidences of fits of all 3 models used to all 4 data sets for the log spiral antenna. The physically motivated models were fit to data with a Gaussian 21cm model added in at a centre frequency of 85MHz. The physically motivated model used $N=9$.}
    \label{tab:Z_comparison}
    \begin{tabular}{|c|c|c|c|c|c|}
        \toprule
           &  & \multicolumn{3}{c}{\centering Model Type}\\
          \cline{3-6}
          & & \multirow{3}{2.8cm}{\centering Polynomial} & \multirow{3}{2.8cm}{\centering Polynomial with Correction} & \multirow{3}{2.8cm}{\centering Physically Motivated with Signal} & \multirow{3}{2.8cm}{\centering Physically Motivated without Signal}\\
          & & & & & \\
          & & & & & \\
          \midrule
           Integration Time & Galaxy Position & \multicolumn{3}{c}{$\log\mathcal{Z}$} \\
         \midrule
         1 Hour & Down & $-483.0 \pm 0.5$ & $-383.5 \pm 0.5$ & $236.7 \pm 0.3$ & $217.6 \pm 0.3$\\
         6 Hours & Down & $-453.2 \pm 0.5$ & $-326.1 \pm 0.6$ & $286.5 \pm 0.3$ & $238.8 \pm 0.3$ \\
         1 Hour & Up & $-869.8 \pm 0.5$ & $-425.0 \pm 0.6$ & $272.1 \pm 0.3$ & $258.8 \pm 0.3$ \\
         6 Hours & Up & $-571.4 \pm 0.5$ & $-389.2 \pm 0.6$ & $276.2 \pm 0.3$ & $241.1 \pm 0.3$\\
         \bottomrule
    \end{tabular}
\end{table*}

The evidences shown in \Cref{tab:Z_comparison} show that the fits using the physically motivated foreground are overwhelmingly the most probable model for all 4 data sets. This demonstrates this physically motivated foreground as a successful means of correcting for antenna chromaticity in these simulated data sets, with the correction being sufficiently accurate to enable a 21cm signal to be identified.

Bayesian evidence also provides a means of determining the confidence with which a signal is detected. The evidence quantifies the probability of a model given the data, marginalised over all parameter values. Therefore, the ratio of the evidences of a joint fit of the signal and foreground model and a fit of the foreground alone will quantify the relative probability of the signal's presence. A significant increase in the evidence when a signal model is included shows a significant statistical preference for the signal being present.

Fits of the physically motivated foreground alone, taking $N=9$ and using the same priors as before, were performed. The evidences are recorded in \Cref{tab:Z_comparison}. In all cases, the log evidence with a signal present is greater than the log evidence without a signal, with the minimum difference being $13.3$ for the data integrated over 1 hour with the galaxy up. A difference of $13.3$ in $\log \mathcal{Z}$ corresponds to betting odds of $\sim600,000:1$ in favour of the higher evidence model. Therefore, all 4 cases strongly statistically favour the presence of a signal.

\subsection{Selecting Model Complexity}\label{sec:model_complexity}
The analysis performed in \Cref{sec:logspiral_results} assumed a foreground model of $N=9$. However, the decision about what value $N$ needs to have, and so how complex the foreground model needs to be, can be derived directly from the data using the Bayesian evidence.

The Bayesian evidence gives a quantification of how likely a given model is based on how well it is able to fit the data. However, it will, by definition, implement Occam's Razor, down-weighting models that have more parameters unless they produce a significant improvement in the fit. Therefore, the value of $N$ required to fit a given data set can be derived by fitting it with models of a range of values of $N$. The model with the highest evidence is then the simplest model capable of producing the best fit to the data.

\begin{figure*}
    \centering
    \includegraphics[width=\textwidth]{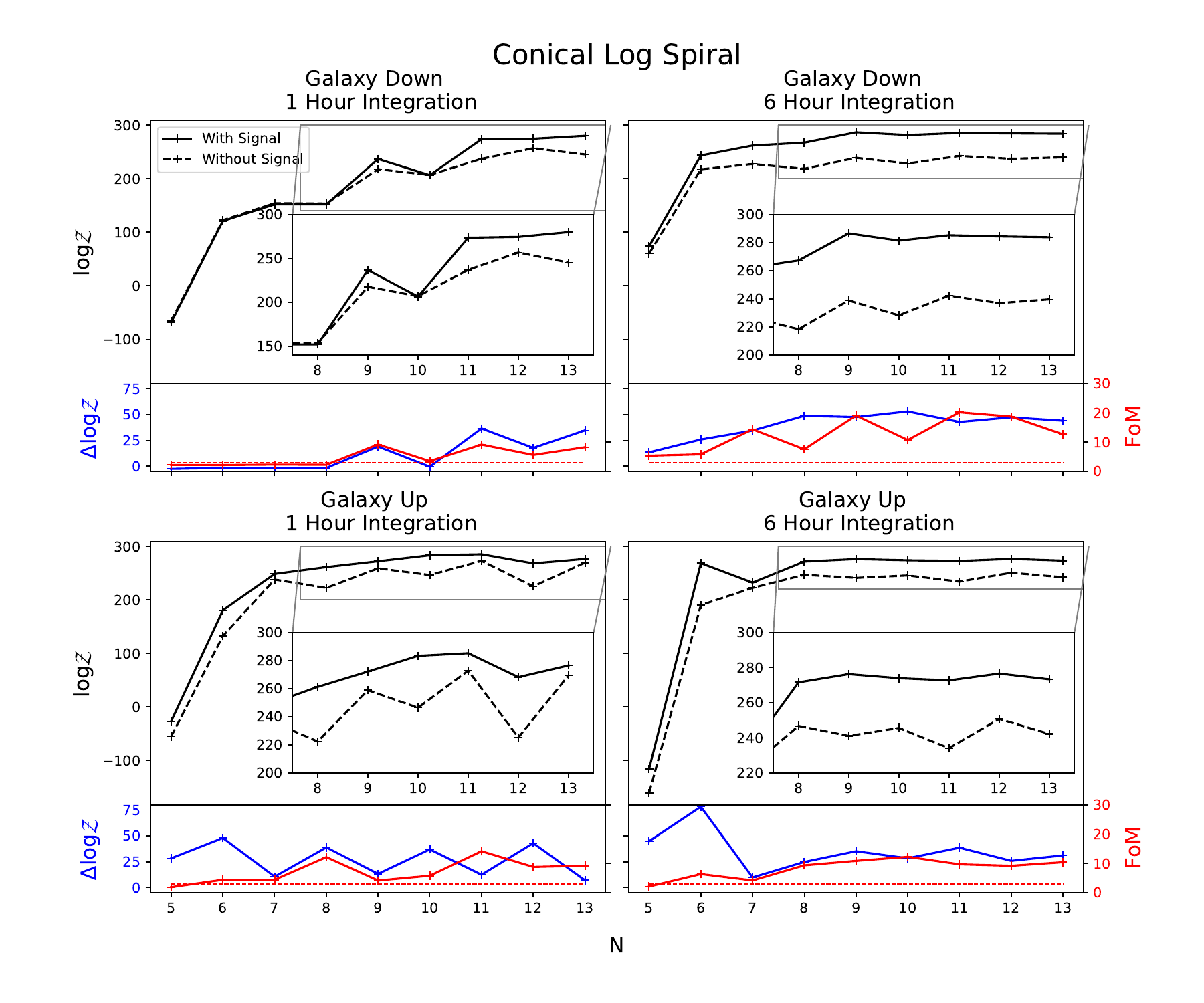}
    \caption{Results of fitting the physically motivated foreground model to simulated data for a log spiral antenna as the complexity of the foreground model varies. In each plot, the upper section shows the absolute $\log \mathcal{Z}$ of a joint fit of the foreground model and a signal model (solid line) and a fit of the foreground model alone (dashed line). A zoomed in view of the highest evidence region is also shown in each case. The lower section shows the difference between the two evidences (blue), together with the Figure of Merit of the signal fit, as defined in \Cref{eq:sig_FoM} (red). The dashed red line shows an approximate FoM threshold for what could be considered a good fit, which is set at 3, above which the rms difference between the true signal and the model is less than $\frac{1}{3}$ of the signal amplitude. A higher $\Delta\log\mathcal{Z}$ indicates a higher degree of confidence in the signal detection, and a higher Figure of Merit indicates a more accurately identified 21cm signal.}
    \label{fig:log_spiral_Z}
\end{figure*}

This was done for the four data sets. \Cref{fig:log_spiral_Z} shows the change in evidence as $N$ increases from 5 to 13. From the Occam's Razor effect intrinsic to Bayesian evidences, it can be expected that the value of $\log\mathcal{Z}$ should rise rapidly from very low values at low $N$, up to a peak, then begin to slowly drop off with little improvement in the quality of the model. This is seen in \Cref{fig:log_spiral_Z}. There are, however, some fluctuations on this pattern of rising to a peak due to way the different numbers of regions divide up structures on the sky and how this interacts with the antenna. $N=13$ was chosen as the cut-off point at which all fits had either peaked or plateaued to where there is no manifest difference between the results of successive $N$ fits beneath the fluctuations.

Statistically significant detection of the signal is maintained for all $N$ above the respective peaks, shown by the high difference in evidence between the models including a signal and those without. It can also be observed that longer integration times perform better than shorter ones, with the signal being detected with a higher degree of confidence and requiring fewer parameters before the evidence levels out. Furthermore, at low integration times, the case with the galaxy above the horizon performs better than the case with the galaxy down, but for higher integration times, there is much less distinction. This is consistent with previous comments.

\Cref{fig:log_spiral_Z} also quantifies the accuracy of the fit signal for each $N$. This is done using a dimensionless Figure of Merit that quantifies the difference between the fit signal and the true signal inserted into the simulated data. This is defined as 

\begin{equation}\label{eq:sig_FoM}
    \mathrm{FoM} = \frac{\mathrm{A}_\mathrm{amp}}{\sqrt{\frac{1}{N_\nu}\sum_{\nu}\left(T_\mathrm{true}\left(\nu\right) - T_\mathrm{fit}\left(\nu, \bar{\theta}\right)\right)^{2}}},
\end{equation}
where $\mathrm{A}_\mathrm{amp}$ is the amplitude of the true signal, 0.155K in this case, $N_\nu$ is the number of frequency data points, $T_\mathrm{true}\left(\nu\right)$ is the true Gaussian signal inserted into the simulated data and $T_\mathrm{fit}\left(\nu, \bar{\theta}\right)$ is the weighted mean fitted signal, where $\bar{\theta}$ refers to the weighted mean of the signal parameters.

\begin{figure*}
    \centering
    \includegraphics[width=\textwidth]{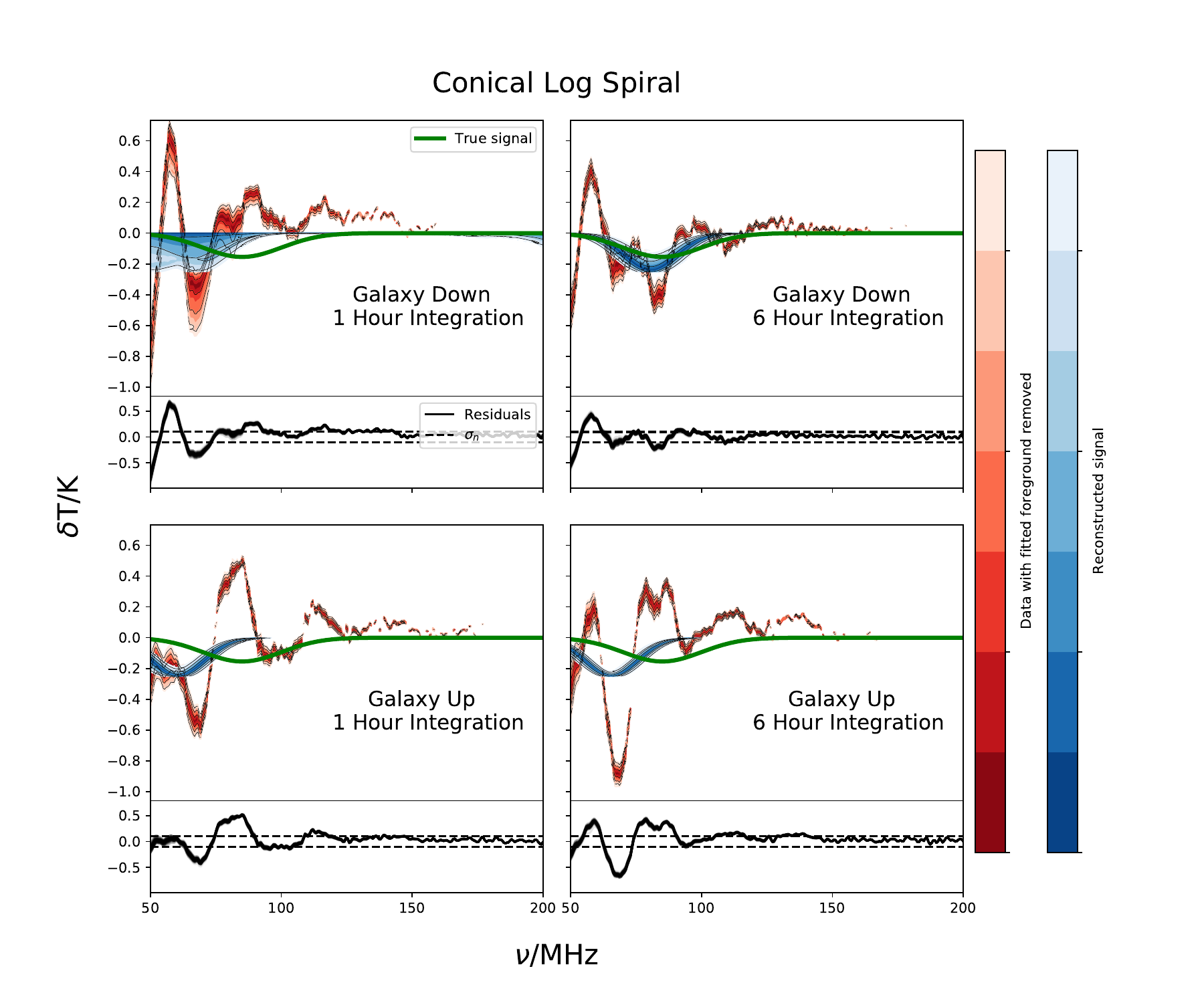}
    \caption{Results from fitting the same data sets as were used in \Cref{fig:new_model_results}, but using a foreground model with $N=5$. Much larger residuals of the foreground model are seen, relative to the $N=9$ case in \Cref{fig:new_model_results}, which mask the 21cm signal.}
    \label{fig:log_spiral_N5}
\end{figure*}

\begin{figure*}
    \centering
    \includegraphics[width=\textwidth]{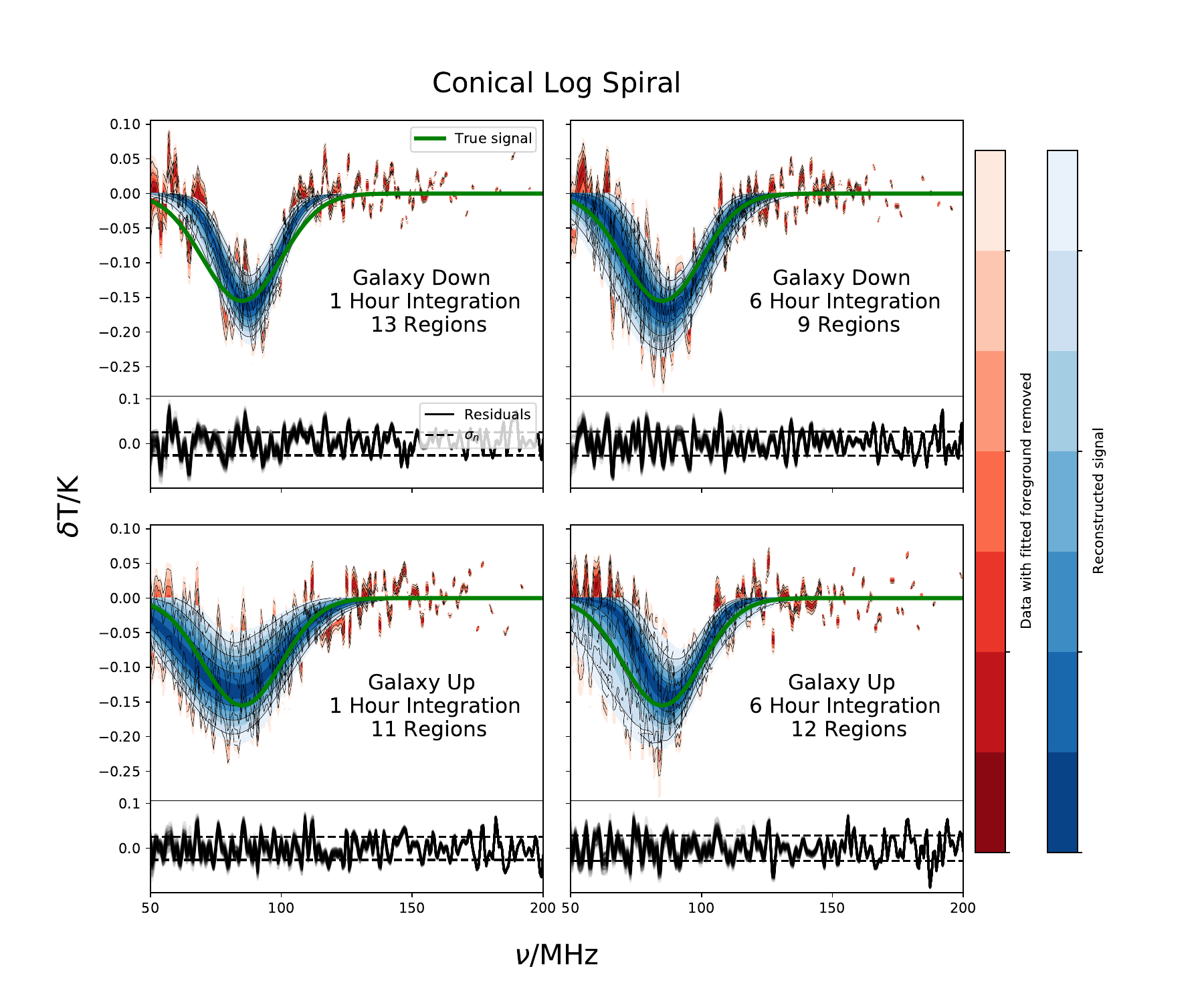}
    \caption{Results from fitting the same data sets as were used in \Cref{fig:new_model_results}, but for the highest evidence $N$ in each case. The 21cm signal is accurately recovered in all 4 cases.}
    \label{fig:log_spiral_N_peak}
\end{figure*}

The importance of using the Bayesian evidence to inform the choice of model complexity in this manner can be demonstrated by considering the results shown in \Cref{fig:log_spiral_N5}, which show the residuals and signal models found when too few regions are used, $N=5$ in this case. Although the residuals are reduced by $\sim2$ orders of magnitude relative to those of a smooth polynomial fit, they are still large enough to mask the 21cm signal. If instead, the Bayesian evidence is used to inform the model, the results in \Cref{fig:log_spiral_N_peak} are found, with the 21cm signal being recovered well in all 4 data sets.

\subsection{Limitations of Approach}\label{sec:limits}
In \Cref{sec:logspiral_results} and \Cref{sec:model_complexity}, we demonstrated that the proposed foreground modelling technique will correct for significant chromatic distortions produced by antenna chromaticity. Provided the model is allowed to be sufficiently detailed, as described by the Bayesian evidence, this enables the 21cm signal to be isolated for the log spiral antenna considered here. However, this technique cannot necessarily correct for any distortions produced by any arbitrarily complex antenna.

One of the other antenna designs we considered was a conical sinuous antenna \citep{buck08}, shown in \Cref{fig:antenna_diagrams}  and \Cref{fig:chromatic_plots}. This has a much more complicated chromatic pattern that the log spiral. The beam patterns shown in \Cref{fig:chromatic_plots} can give a measure of the chromaticity of the two antennae. The log spiral antenna can be seen to have approximately constant directivity in any given direction over the entire frequency band. There is some oscillation in directivity around the central value of each pixel, with the amplitude of oscillation dropping off away from zenith, together with the magnitude of the oscillation centre. Furthermore, all directions have oscillations in directivity that are approximately in phase. The conical sinuous antenna, however, has a much more complex pattern. The variations in pixel directivity can be more than twice as large as those of the log spiral and have less correlation between the variations of adjacent pixels.

To characterise the effects of this increased chromaticity, the tests performed in \Cref{sec:logspiral_results} and \Cref{sec:model_complexity} on the log spiral antenna were performed for the conical sinuous antenna.

\begin{figure*}
    \centering
    \includegraphics[width=\textwidth]{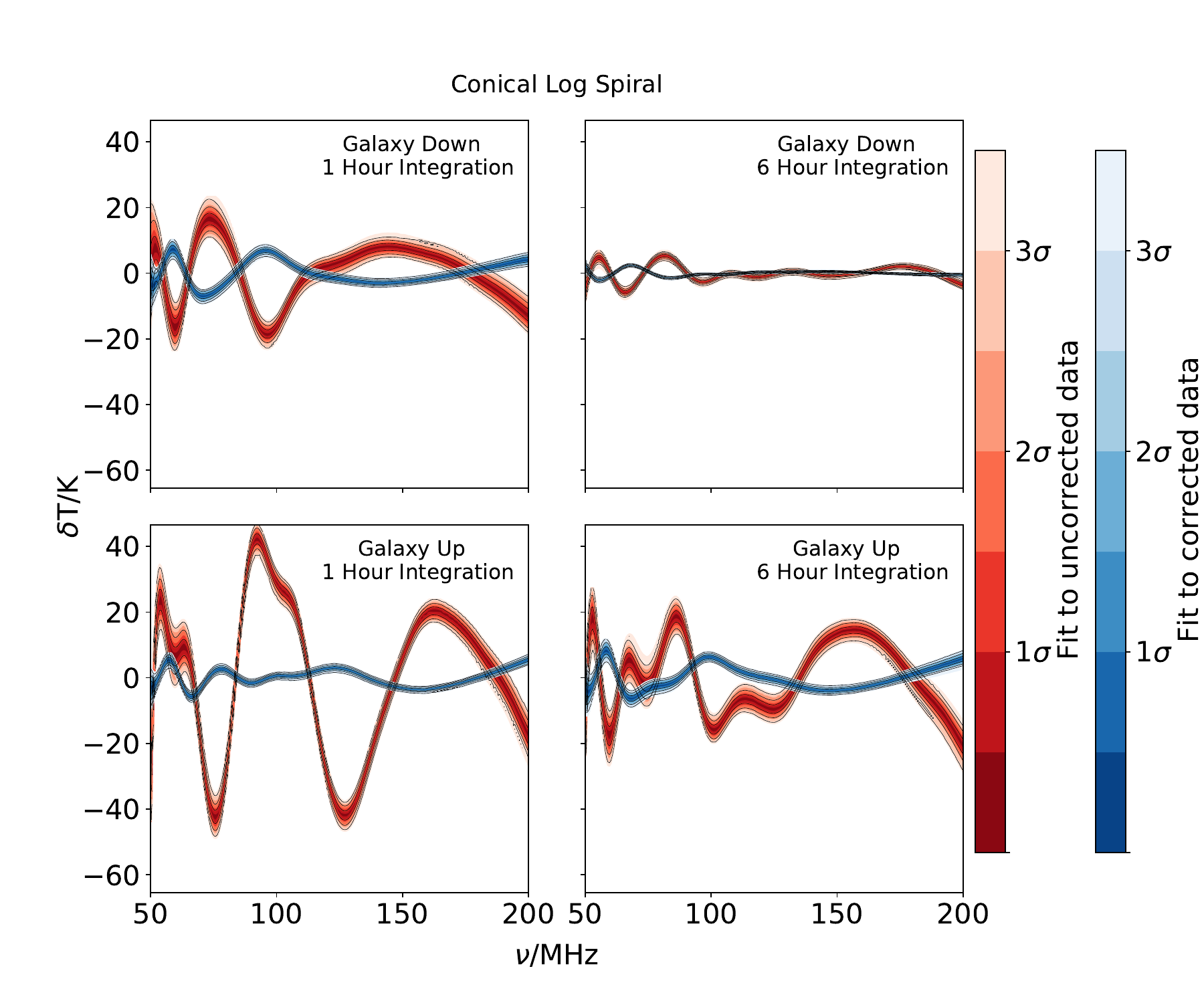}
    \caption{Residuals after subtraction of the best fit $5^\mathrm{th}$ order log polynomial from corrected (red) and uncorrected (blue) simulated observation data using a conical sinuous antenna. Observation parameters are the same as those in \Cref{fig:polyfit_residuals}.}
    \label{fig:polyfit_residuals_con_sin}
\end{figure*}

Fitting a simulated data set generated with this antenna with a $5^\mathrm{th}$ order log polynomial produces the residuals shown in \Cref{fig:polyfit_residuals_con_sin}. These residuals are of similar magnitude to those of the log spiral antenna. Likewise, the results of a polynomial foreground fit after performing the uniform $\beta$ chromaticity correction described in \Cref{sec:simple_correction} are also shown. These also demonstrate a reduction of the residuals, but not enough to allow the 21cm signal to be identified, as was seen in the log spiral. 

\begin{figure*}
    \centering
    \includegraphics[width=\textwidth]{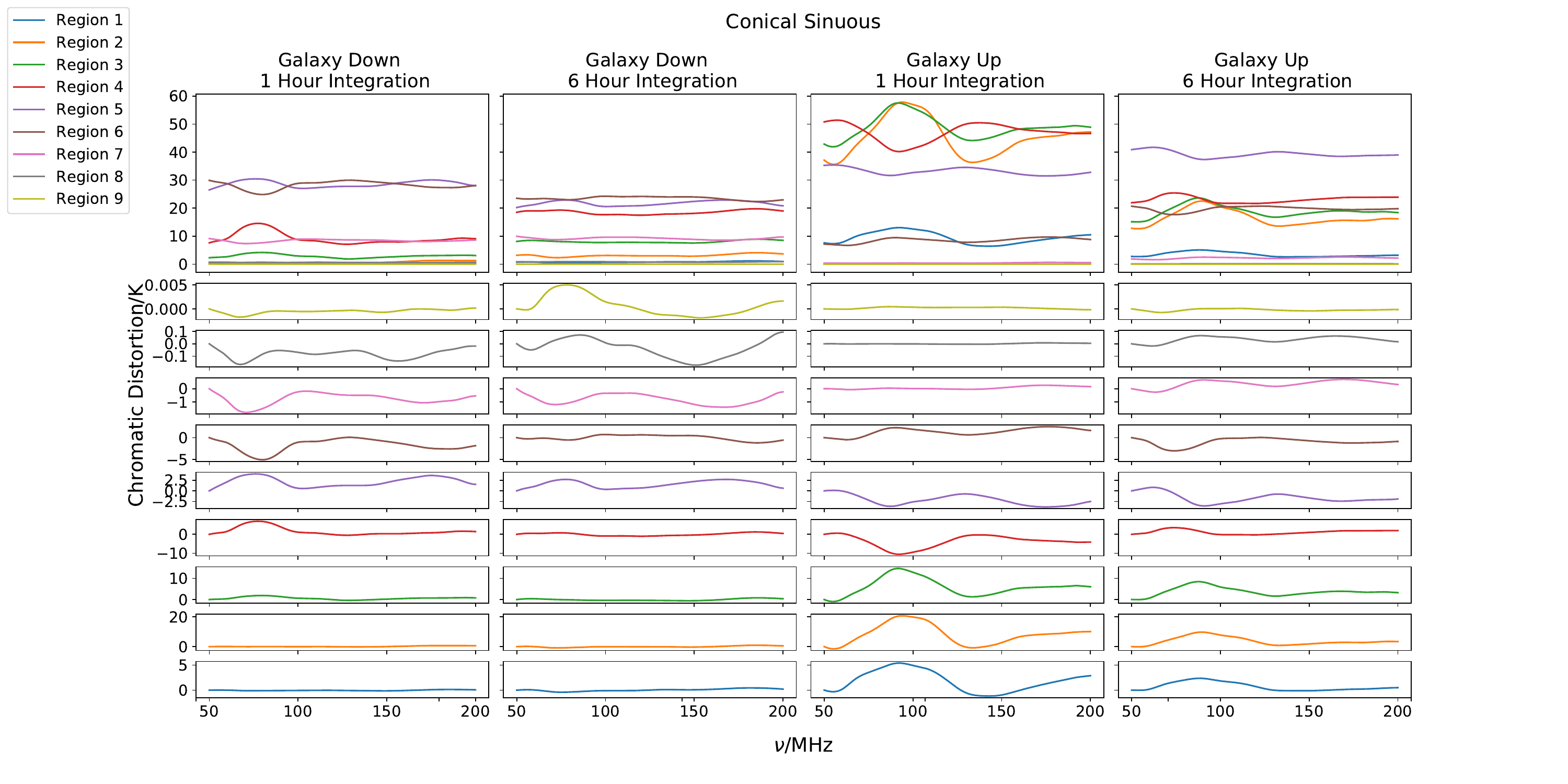}
    \caption{Chromaticity functions for a conical sinuous antenna when $N=9$. Layout is the same as in \Cref{fig:log_spir_chromatic_functions}}
    \label{fig:con_sin_chromatic_functions}
\end{figure*}

The chromaticity functions for the conical sinuous antenna for $N=9$ are shown in \Cref{fig:con_sin_chromatic_functions}. The functions have fluctuations of much greater magnitude than the log spiral antenna. They are also much less regular across the band, as compared to the near sinusoidal chromaticity functions of the log spiral.

\begin{figure*}
    \centering
    \includegraphics[width=\textwidth]{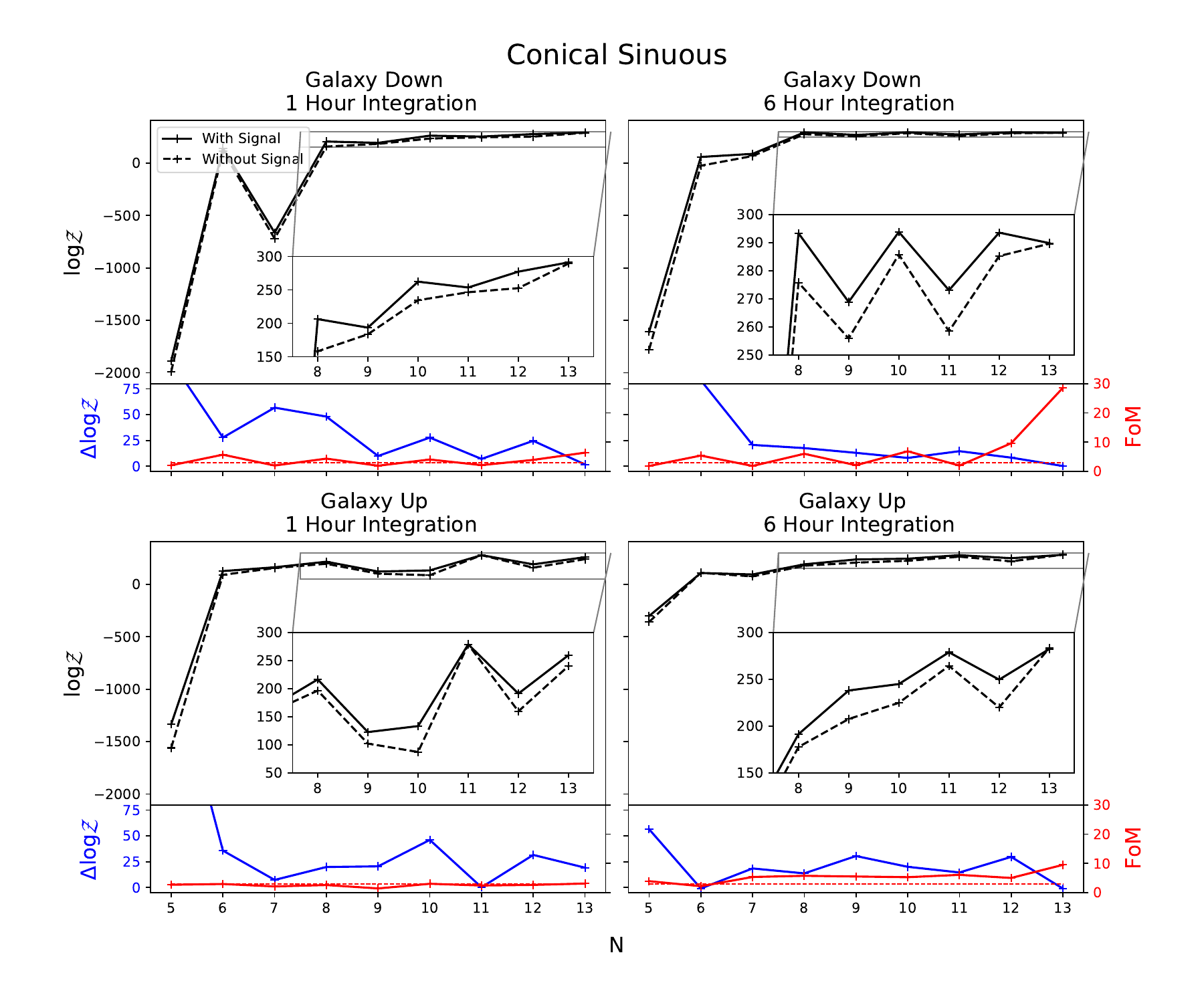}
    \caption{Equivalent of \Cref{fig:log_spiral_Z}, but for a conical sinuous antenna}
    \label{fig:conical_sinuous_Z}
\end{figure*}

The conical sinuous simulated data sets, with Gaussian 21cm signals added, were also fit using the physically motivated foreground model for various values of $N$. \Cref{fig:conical_sinuous_Z} summarises the evidences of these fits for models of the foreground and a Gaussian signal, and for the foreground alone. The same effect as was observed for the log spiral antenna can be seen here, with the evidence rapidly dropping off at lower $N$, as the complexity of the foreground model becomes insufficient to fit the structure present, and reaching a peak at higher $N$.

\begin{figure*}
    \centering
    \includegraphics[width=\textwidth]{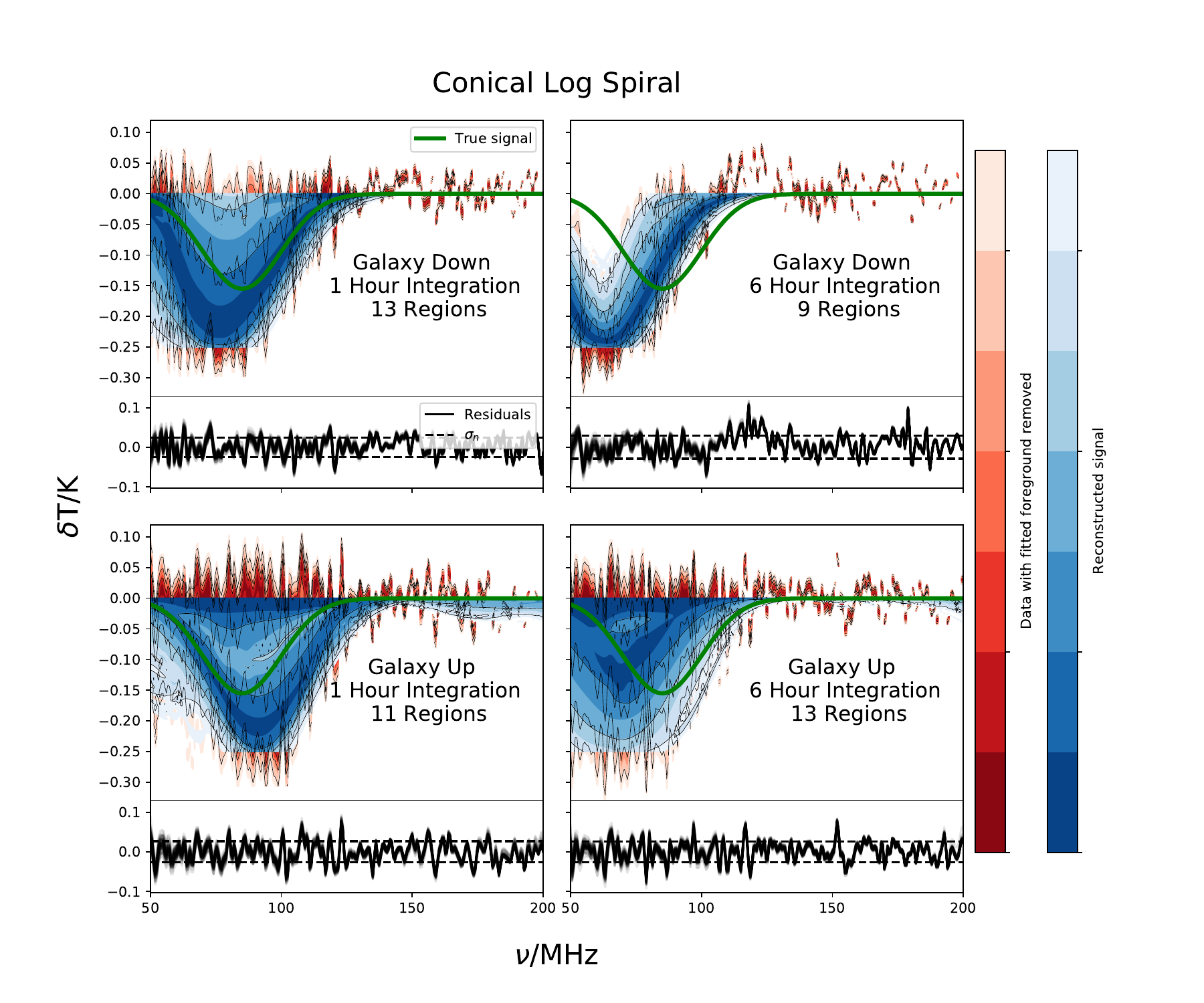}
    \caption{Equivalent to \Cref{fig:log_spiral_N_peak}, but for a conical sinuous antenna}
    \label{fig:con_sin_N_peak}
\end{figure*}

It can be seen from \Cref{fig:conical_sinuous_Z}, that the fits using this conical sinuous antenna have lower $\Delta\log\mathcal{Z}$, and so less confidence in the signal detections at the peak evidence values, than fits using the log spiral antenna. This is indicative that the larger chromatic distortions of the conical sinuous antenna are more degenerate with the 21cm signal and harder to distinguish. This is demonstrated by \Cref{fig:con_sin_N_peak}, which shows the residuals and signal models for the highest evidence $N$ fits using the conical sinuous antenna, which were 13, 9, 11 and 13 for 1 hour integration with the galaxy down, 6 hours integration with the galaxy down, 1 hour integration with the galaxy up and 6 hours integration with the galaxy up, respectively. The low confidences of these detections demonstrate that if the antenna being used is too strongly chromatic, this physically motivated foreground model may be insufficient to allow the 21cm signal to be isolated.

\begin{figure}
    \centering
    \includegraphics[width=\columnwidth]{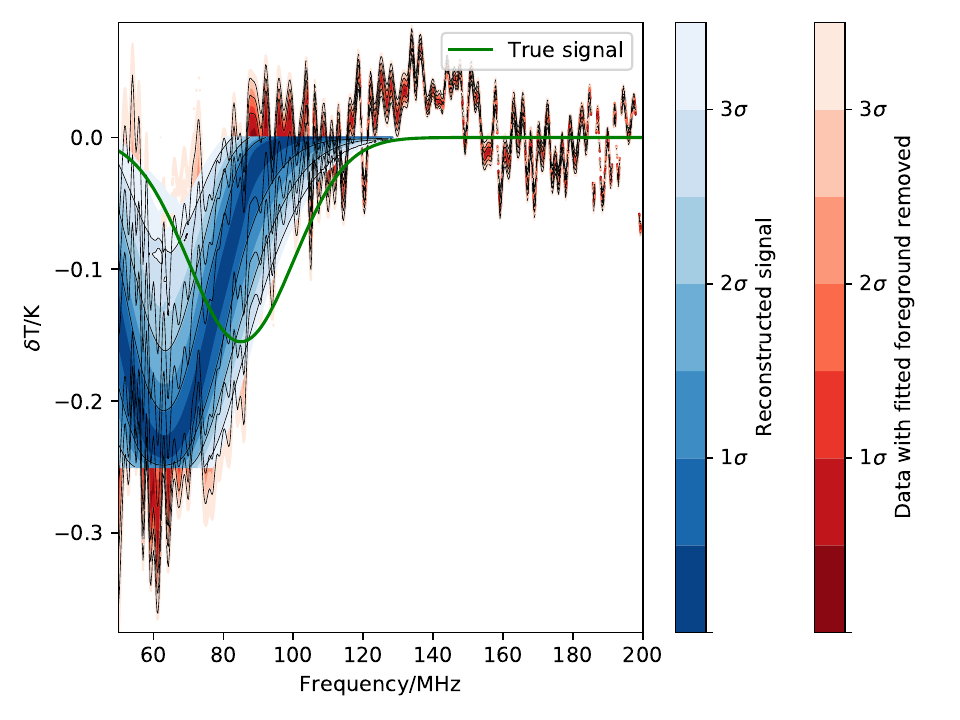}
    \caption{Plot of the results of fitting simulated data with 6 hours of integration and the galaxy down for a conical sinuous antenna, using a foreground model with $N=11$.}
    \label{fig:con_sin_false_signal}
\end{figure}

Another effect that can be seen in the results of testing this physically motivated foreground fitting using a highly chromatic conical sinuous antenna, is that there are some results with a high $\log\mathcal{Z}$, at or near the evidence peak, that show a fit with a high $\Delta\log\mathcal{Z}$ confidence in the signal found, but a low Figure of Merit. This was not observed with the log spiral antenna. This corresponds to a systematic error in which a false, incorrect signal is being confidently detected in the data. For example, this can be seen in the case of $N=11$, for 6 hours of integration with the galaxy down. The corresponding residuals are given in \Cref{fig:con_sin_false_signal}.

\begin{figure}
    \centering
    \includegraphics[width=\columnwidth]{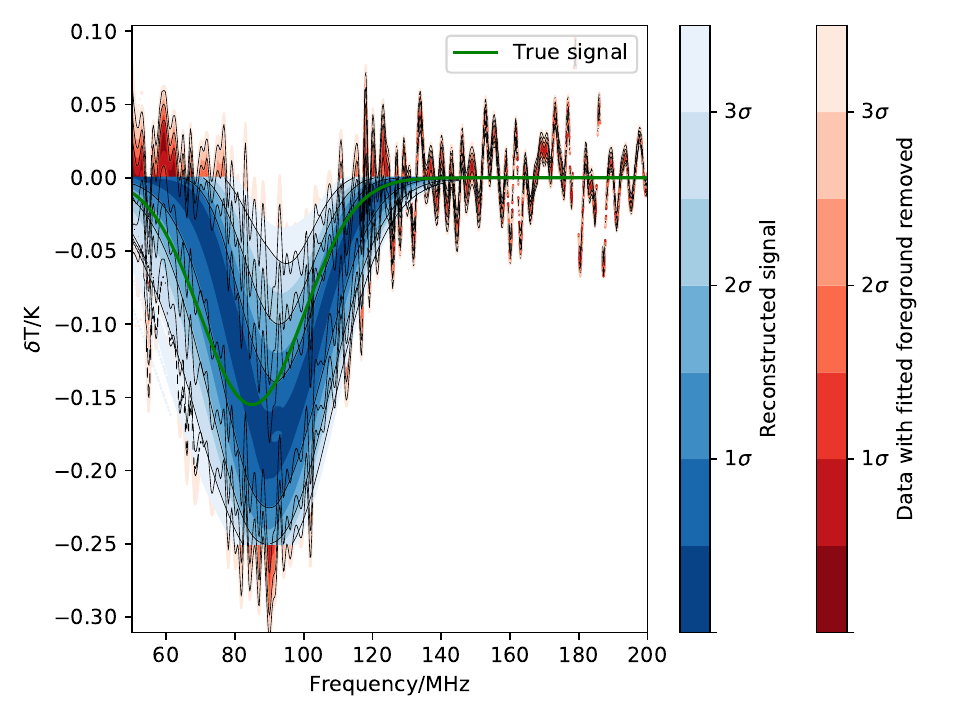}
    \caption{Plot of the results of fitting simulated data with 6 hours of integration and the galaxy down for a conical sinuous antenna, using a foreground model with $N=10$.}
    \label{fig:con_sin_true_signal}
\end{figure}

This demonstrates the possibility of highly chromatic antennae producing a false confident detection of the 21cm signal when a highly chromatic antenna is used. However, such systematic errors can be identified when using this foreground modelling method. This is done by considering model fits at many different $N$ values. In the case of the log spiral antenna, the 21cm signal models found by model fits using successive $N$ values around the highest evidence $N$ are in agreement with each other. Each fit detects the same signal. However, in the case of the conical sinuous, the signal found can be observed to change with successive $N$ values, as can be seen in \Cref{fig:con_sin_true_signal}. This helps identify that such signals are still subject to systematic distortions. 

Overall, the results of the conical sinuous antenna demonstrate that, whilst the foreground model described here can successfully correct for chromatic distortions in global 21cm experiment data sufficiently to identify a 21cm signal, it will not necessarily be able to do so for any arbitrarily chromatic antenna. The method was successful in enabling a 21cm signal to be accurately identified using a log spiral antenna, despite it being sufficiently chromatic as to mask the 21cm signal if no correction, or even a simple correction assuming an unchanging power distribution, is made. However, for the much more strongly chromatic conical sinuous antenna, a 21cm signal injected at 85MHz could not be identified, even as $N$ was made large. In practice, the model complexity $N$ could be increased further. However, as the value of $N$ at which the evidence peaks makes for a good indication of the quality of the antenna and general setup, such as integration time, this is sufficient for comparison with the log spiral antennae, which peaks in evidence at much lower $N$.

A further limitation of this method of fitting physically motivated foregrounds arises due to the noise model used. The simulated data used in this paper were all generated with uncorrelated Gaussian noise, and the likelihoods used also all assumed uncorrelated Gaussian noise. In practice, it is not reasonable to assume that the noise of each channel of data from a global 21cm experiment would be uncorrelated. This will lead to the likelihoods used in this paper having more statistical information than would be the case when fitting real data. The more realistic case of using a physically motivated, realistic noise model will be considered in future work.

\section{Conclusions}\label{sec:conclusions}
The ability to detect the HI 21cm signal from beneath the galactic and extra-galactic foregrounds is heavily dependent on the ability to accurately model systematic distortions in the data. A significant proportion of these systematics arise from the coupling of a chromatic antenna with the complicated, highly non-uniform spatial distribution of foreground power across the sky.

We have demonstrated that, for an antenna that is not perfectly achromatic, such as a log spiral antenna, the chromatic distortions are several orders of magnitude greater than the expected 21cm signal amplitude. This excess of distortion would make identifying the 21cm signal in the data with a simple, smooth foreground model nearly impossible. This was found to be true even with the galaxy below the horizon and with increasing integration time. We also found that applying a simple chromatic correction that assumes an unchanging temperature distribution on the sky with changing frequency, apart from a uniform scaling by a power law, was unable to overcome the distortions for this antenna. The change in foreground spatial distribution from a non-uniform spectral index was sufficient to inhibit the correction. However, very little data exists about the spectral index distribution at low frequencies, meaning this effect cannot be accounted for via a chromaticity correction factor calculated from previously known data.

We proposed a new foreground model designed to use physical simulations of the foregrounds and antenna pattern in the model fitting itself, in order to fit the chromatic distortions as part of the foreground. This foreground model works by dividing the sky into predefined regions in which the spectral index is expected to be similar. A foreground function is then produced by scaling a base all-sky map to each observing frequency by assuming each region has a uniform spectral index and curvature, and convolving the result with a model of the beam. The spectral indices and curvatures of each region can then be fitted for, using the data.

This model is also designed to have adjustable complexity. By increasing the number of regions the sky is divided into, the model becomes more detailed and able to more closely match the true sky, at the cost of more parameters. By fitting this model in a Bayesian sense using a nested sampling algorithm, the Bayesian evidence can be used to identify how complex the model needs to be directly from the data.

We tested how this new foreground model performed on simulated data of a log spiral antenna. This was found to be a great improvement on the results of a simple chromatic correction or no correction at all. The new model was able to model the chromatic distortions of the log spiral antenna to sufficiently high accuracy as to enable a simulated Gaussian 21cm signal at 85MHz to be confidently and accurately detected in the data. For real data, a more detailed physical model of the 21cm signal would be necessary, but the Gaussian was adequate here to demonstrate whether a detection would be possible.

Overall, these results demonstrate that the proposed model is capable of correcting for a significant amount of chromatic distortion in global 21cm data. The model enabled the detection of a 21cm signal in data simulated using an antenna that was sufficiently strongly chromatic that assuming smooth foregrounds or making a simple chromaticity correction was inadequate.

This detection is also achieved even when the model for chromatic distortion is averaged over the entire period of observation. If the model was instead fit to the data at many different LSTs simultaneously, thus exploiting a greater amount of information present in the chromatic structure, the performance of this method would, in theory, improve further. This will be explored in future work.

However, the proposed model was not found to be able to correct for any arbitrarily chromatic antenna distortions. Repeating the tests on a much more strongly chromatic conical sinuous antenna found that it did not perform as well. Although the physically motivated model gave a much better fit to the data than a smooth polynomial foreground model, it was found to be insufficient to allow a 21cm signal at $85\mathrm{MHz}$ to be confidently detected. This is possibly due to the structures possible in the foreground model being too strongly degenerate with the 21cm signal. This means that it could not detect the signal to a high degree of confidence, even as the model was made highly complex. This means that simulated observations and analysis using this technique can be used to determine if an antenna is too strongly chromatic to allow detection of the 21cm signal, and so can be used to guide the antenna design of a global 21cm experiment. This process will be demonstrated in a subsequent work.

In general, however, the proposed foreground model was found to enable the detection of a 21cm signal with moderately chromatic antennae such as a log spiral, even when that antenna is too strongly chromatic to allow a detection of the global 21cm signal when using conventional data analysis techniques.

\section*{Acknowledgements}

Dominic Anstey and Eloy de Lera Acedo thank the Science and Technologies Facilities Council and Will Handley thanks Gonville and Caius College for their support. We would also like to thank the Kavli foundation etc. for their support of the REACH experiment. Furthermore, we thank John Cumner and Quentin Gueuning for providing the electromagnetic simulations of the antennae used in this work.

\section*{Data Availability}
The data underlying this article is available on Zenodo, at https://dx.doi.org/10.5281/zenodo.4599746.

%%%%%%%%%%%%%%%%%%%%%%%%%%%%%%%%%%%%%%%%%%%%%%%%%%



\bibliographystyle{mnras}
\bibliography{bib} % if your bibtex file is called example.bib

\begin{thebibliography}{}
\makeatletter
\relax
\def\mn@urlcharsother{\let\do\@makeother \do\$\do\&\do\#\do\^\do\_\do\%\do\~}
\def\mn@doi{\begingroup\mn@urlcharsother \@ifnextchar [ {\mn@doi@}
  {\mn@doi@[]}}
\def\mn@doi@[#1]#2{\def\@tempa{#1}\ifx\@tempa\@empty \href
  {http://dx.doi.org/#2} {doi:#2}\else \href {http://dx.doi.org/#2} {#1}\fi
  \endgroup}
\def\mn@eprint#1#2{\mn@eprint@#1:#2::\@nil}
\def\mn@eprint@arXiv#1{\href {http://arxiv.org/abs/#1} {{\tt arXiv:#1}}}
\def\mn@eprint@dblp#1{\href {http://dblp.uni-trier.de/rec/bibtex/#1.xml}
  {dblp:#1}}
\def\mn@eprint@#1:#2:#3:#4\@nil{\def\@tempa {#1}\def\@tempb {#2}\def\@tempc
  {#3}\ifx \@tempc \@empty \let \@tempc \@tempb \let \@tempb \@tempa \fi \ifx
  \@tempb \@empty \def\@tempb {arXiv}\fi \@ifundefined
  {mn@eprint@\@tempb}{\@tempb:\@tempc}{\expandafter \expandafter \csname
  mn@eprint@\@tempb\endcsname \expandafter{\@tempc}}}

\bibitem[\protect\citeauthoryear{{Barkana}}{{Barkana}}{2018}]{barkana18b}
{Barkana} R.,  2018, {Possible interaction between baryons and dark-matter
  particles revealed by the first stars} (\mn@eprint {arXiv} {1803.06698}),
  \mn@doi{10.1038/nature25791}

\bibitem[\protect\citeauthoryear{{Barkana}, {Outmezguine}, {Redigol}  \&
  {Volansky}}{{Barkana} et~al.}{2018}]{barkana18a}
{Barkana} R.,  {Outmezguine} N.~J.,  {Redigol} D.,   {Volansky} T.,  2018,
  {Strong constraints on light dark matter interpretation of the EDGES signal}
  (\mn@eprint {arXiv} {1803.03091}), \mn@doi{10.1103/PhysRevD.98.103005}

\bibitem[\protect\citeauthoryear{Berlin, Hooper, Krnjaic  \& McDermott}{Berlin
  et~al.}{2018}]{berlin18}
Berlin A.,  Hooper D.,  Krnjaic G.,   McDermott S.~D.,  2018, Severely
  Constraining Dark-Matter Interpretations of the 21-cm Anomaly,
  \mn@doi{10.1103/PhysRevLett.121.011102}, \url
  {https://link.aps.org/doi/10.1103/PhysRevLett.121.011102}

\bibitem[\protect\citeauthoryear{{Bernardi} et~al.,}{{Bernardi}
  et~al.}{2016}]{bernardi16}
{Bernardi} G.,  et~al., 2016, \mn@doi [\mnras] {10.1093/mnras/stw1499}, \href
  {https://ui.adsabs.harvard.edu/abs/2016MNRAS.461.2847B} {461, 2847}

\bibitem[\protect\citeauthoryear{{Bevins}, {Handley}, {Fialkov}, {de Lera
  Acedo}, {Greenhill}  \& {Price}}{{Bevins} et~al.}{2020}]{bevins20}
{Bevins} H.~T.~J.,  {Handley} W.~J.,  {Fialkov} A.,  {de Lera Acedo} E.,
  {Greenhill} L.~J.,   {Price} D.~C.,  2020, arXiv e-prints, \href
  {https://ui.adsabs.harvard.edu/abs/2020arXiv200714970B} {p. arXiv:2007.14970}

\bibitem[\protect\citeauthoryear{{Bowman}, {Rogers}  \& {Hewitt}}{{Bowman}
  et~al.}{2008}]{bowman08}
{Bowman} J.~D.,  {Rogers} A. E.~E.,   {Hewitt} J.~N.,  2008, \mn@doi [\apj]
  {10.1086/528675}, \href
  {https://ui.adsabs.harvard.edu/abs/2008ApJ...676....1B} {676, 1}

\bibitem[\protect\citeauthoryear{{Bowman}, {Rogers}, {Monsalve}, {Mozdzen}  \&
  {Mahesh}}{{Bowman} et~al.}{2018}]{bowman18}
{Bowman} J.~D.,  {Rogers} A. E.~E.,  {Monsalve} R.~A.,  {Mozdzen} T.~J.,
  {Mahesh} N.,  2018, \mn@doi [\nat] {10.1038/nature25792}, \href
  {https://ui.adsabs.harvard.edu/abs/2018Natur.555...67B} {555, 67}

\bibitem[\protect\citeauthoryear{{Buck} \& {Filipovic}}{{Buck} \&
  {Filipovic}}{2008}]{buck08}
{Buck} M.~C.,  {Filipovic} D.~S.,  2008, \mn@doi [IEEE Transactions on Antennas
  and Propagation] {10.1109/TAP.2008.922606}, \href
  {https://ui.adsabs.harvard.edu/abs/2008ITAP...56.1229B} {56, 1229}

\bibitem[\protect\citeauthoryear{Chib \& Greenberg}{Chib \&
  Greenberg}{1995}]{chib95}
Chib S.,  Greenberg E.,  1995, \mn@doi [The American Statistician]
  {10.1080/00031305.1995.10476177}, 49, 327

\bibitem[\protect\citeauthoryear{{Cohen}, {Fialkov}, {Barkana}  \&
  {Lotem}}{{Cohen} et~al.}{2017}]{cohen17}
{Cohen} A.,  {Fialkov} A.,  {Barkana} R.,   {Lotem} M.,  2017, {Charting the
  parameter space of the global 21-cm signal} (\mn@eprint {arXiv}
  {1609.02312}), \mn@doi{10.1093/mnras/stx2065}

\bibitem[\protect\citeauthoryear{{Cohen}, {Fialkov}, {Barkana}  \&
  {Monsalve}}{{Cohen} et~al.}{2020}]{cohen20}
{Cohen} A.,  {Fialkov} A.,  {Barkana} R.,   {Monsalve} R.~A.,  2020, \mn@doi
  [\mnras] {10.1093/mnras/staa1530}, \href
  {https://ui.adsabs.harvard.edu/abs/2020MNRAS.495.4845C} {495, 4845}

\bibitem[\protect\citeauthoryear{{DeBoer} et~al.,}{{DeBoer}
  et~al.}{2017}]{deboer17}
{DeBoer} D.~R.,  et~al., 2017, \mn@doi [\pasp]
  {10.1088/1538-3873/129/974/045001}, \href
  {https://ui.adsabs.harvard.edu/abs/2017PASP..129d5001D} {129, 045001}

\bibitem[\protect\citeauthoryear{{Dewdney}, {Hall}, {Schilizzi}  \&
  {Lazio}}{{Dewdney} et~al.}{2009}]{dewdney09}
{Dewdney} P.~E.,  {Hall} P.~J.,  {Schilizzi} R.~T.,   {Lazio} T.~J.~L.~W.,
  2009, \mn@doi [IEEE Proceedings] {10.1109/JPROC.2009.2021005}, \href
  {https://ui.adsabs.harvard.edu/abs/2009IEEEP..97.1482D} {97, 1482}

\bibitem[\protect\citeauthoryear{{Dyson}}{{Dyson}}{1965}]{dyson65}
{Dyson} J.,  1965, \mn@doi [IEEE Transactions on Antennas and Propagation]
  {10.1109/TAP.1965.1138471}, \href
  {https://ui.adsabs.harvard.edu/abs/1965ITAP...13..488D} {13, 488}

\bibitem[\protect\citeauthoryear{{Ewall-Wice}, {Chang}, {Lazio}, {Dor{\'e}},
  {Seiffert}  \& {Monsalve}}{{Ewall-Wice} et~al.}{2018}]{ewall18}
{Ewall-Wice} A.,  {Chang} T.~C.,  {Lazio} J.,  {Dor{\'e}} O.,  {Seiffert} M.,
  {Monsalve} R.~A.,  2018, {Modeling the Radio Background from the First Black
  Holes at Cosmic Dawn: Implications for the 21 cm Absorption Amplitude}
  (\mn@eprint {arXiv} {1803.01815}), \mn@doi{10.3847/1538-4357/aae51d}

\bibitem[\protect\citeauthoryear{{Ewall-Wice}, {Chang}  \&
  {Lazio}}{{Ewall-Wice} et~al.}{2020}]{ewall20}
{Ewall-Wice} A.,  {Chang} T.-C.,   {Lazio} T. J.~W.,  2020, {The Radio Scream
  from black holes at Cosmic Dawn: a semi-analytic model for the impact of
  radio-loud black holes on the 21 cm global signal} (\mn@eprint {arXiv}
  {1903.06788}), \mn@doi{10.1093/mnras/stz3501}

\bibitem[\protect\citeauthoryear{{Feng} \& {Holder}}{{Feng} \&
  {Holder}}{2018}]{feng18}
{Feng} C.,  {Holder} G.,  2018, {Enhanced Global Signal of Neutral Hydrogen Due
  to Excess Radiation at Cosmic Dawn} (\mn@eprint {arXiv} {1802.07432}),
  \mn@doi{10.3847/2041-8213/aac0fe}

\bibitem[\protect\citeauthoryear{{Feroz}, {Hobson}  \& {Bridges}}{{Feroz}
  et~al.}{2009}]{feroz09}
{Feroz} F.,  {Hobson} M.~P.,   {Bridges} M.,  2009, \mn@doi [\mnras]
  {10.1111/j.1365-2966.2009.14548.x}, \href
  {https://ui.adsabs.harvard.edu/abs/2009MNRAS.398.1601F} {398, 1601}

\bibitem[\protect\citeauthoryear{{Fialkov} \& {Barkana}}{{Fialkov} \&
  {Barkana}}{2019}]{fialkov19}
{Fialkov} A.,  {Barkana} R.,  2019, {Signature of excess radio background in
  the 21-cm global signal and power spectrum} (\mn@eprint {arXiv}
  {1902.02438}), \mn@doi{10.1093/mnras/stz873}

\bibitem[\protect\citeauthoryear{{Field}}{{Field}}{1958}]{field58}
{Field} G.~B.,  1958, \mn@doi [Proceedings of the IRE]
  {10.1109/JRPROC.1958.286741}, \href
  {https://ui.adsabs.harvard.edu/abs/1958PIRE...46..240F} {46, 240}

\bibitem[\protect\citeauthoryear{{Foreman-Mackey}, {Hogg}, {Lang}  \&
  {Goodman}}{{Foreman-Mackey} et~al.}{2013}]{foremanmackey13}
{Foreman-Mackey} D.,  {Hogg} D.~W.,  {Lang} D.,   {Goodman} J.,  2013, \mn@doi
  [\pasp] {10.1086/670067}, \href
  {https://ui.adsabs.harvard.edu/abs/2013PASP..125..306F} {125, 306}

\bibitem[\protect\citeauthoryear{{Franx}, {Illingworth}, {Kelson}, {van Dokkum}
   \& {Tran}}{{Franx} et~al.}{1997}]{franx97}
{Franx} M.,  {Illingworth} G.~D.,  {Kelson} D.~D.,  {van Dokkum} P.~G.,
  {Tran} K.-V.,  1997, \mn@doi [\apj] {10.1086/310844}, \href
  {https://ui.adsabs.harvard.edu/abs/1997ApJ...486L..75F} {486, L75}

\bibitem[\protect\citeauthoryear{{Furlanetto}}{{Furlanetto}}{2016}]{furlanetto16}
{Furlanetto} S.~R.,  2016, in {Mesinger} A.,  ed.,  Astrophysics and Space
  Science Library Vol. 423, Understanding the Epoch of Cosmic Reionization:
  Challenges and Progress. p.~247 (\mn@eprint {arXiv} {1511.01131}),
  \mn@doi{10.1007/978-3-319-21957-8_9}

\bibitem[\protect\citeauthoryear{{Handley}, {Hobson}  \& {Lasenby}}{{Handley}
  et~al.}{2015a}]{handley15b}
{Handley} W.~J.,  {Hobson} M.~P.,   {Lasenby} A.~N.,  2015a, \mn@doi [\mnras]
  {10.1093/mnrasl/slv047}, \href
  {https://ui.adsabs.harvard.edu/abs/2015MNRAS.450L..61H} {450, L61}

\bibitem[\protect\citeauthoryear{{Handley}, {Hobson}  \& {Lasenby}}{{Handley}
  et~al.}{2015b}]{handley15a}
{Handley} W.~J.,  {Hobson} M.~P.,   {Lasenby} A.~N.,  2015b, \mn@doi [\mnras]
  {10.1093/mnras/stv1911}, \href
  {https://ui.adsabs.harvard.edu/abs/2015MNRAS.453.4384H} {453, 4384}

\bibitem[\protect\citeauthoryear{{Hills}, {Kulkarni}, {Meerburg}  \&
  {Puchwein}}{{Hills} et~al.}{2018}]{hills18}
{Hills} R.,  {Kulkarni} G.,  {Meerburg} P.~D.,   {Puchwein} E.,  2018, \mn@doi
  [\nat] {10.1038/s41586-018-0796-5}, \href
  {https://ui.adsabs.harvard.edu/abs/2018Natur.564E..32H} {564, E32}

\bibitem[\protect\citeauthoryear{{Knox}, {Scoccimarro}  \& {Dodelson}}{{Knox}
  et~al.}{1998}]{knox98}
{Knox} L.,  {Scoccimarro} R.,   {Dodelson} S.,  1998, \mn@doi [\prl]
  {10.1103/PhysRevLett.81.2004}, \href
  {https://ui.adsabs.harvard.edu/abs/1998PhRvL..81.2004K} {81, 2004}

\bibitem[\protect\citeauthoryear{Liu, Outmezguine, Redigolo  \& Volansky}{Liu
  et~al.}{2019}]{liu19}
Liu H.,  Outmezguine N.~J.,  Redigolo D.,   Volansky T.,  2019, Reviving
  millicharged dark matter for 21-cm cosmology,
  \mn@doi{10.1103/PhysRevD.100.123011}, \url
  {https://link.aps.org/doi/10.1103/PhysRevD.100.123011}

\bibitem[\protect\citeauthoryear{{Lonsdale} et~al.,}{{Lonsdale}
  et~al.}{2009}]{lonsdale09}
{Lonsdale} C.~J.,  et~al., 2009, \mn@doi [IEEE Proceedings]
  {10.1109/JPROC.2009.2017564}, \href
  {https://ui.adsabs.harvard.edu/abs/2009IEEEP..97.1497L} {97, 1497}

\bibitem[\protect\citeauthoryear{{Mirocha} \& {Furlanetto}}{{Mirocha} \&
  {Furlanetto}}{2019}]{mirocha19}
{Mirocha} J.,  {Furlanetto} S.~R.,  2019, {What does the first highly
  redshifted 21-cm detection tell us about early galaxies?} (\mn@eprint {arXiv}
  {1803.03272}), \mn@doi{10.1093/mnras/sty3260}

\bibitem[\protect\citeauthoryear{{Monsalve}, {Fialkov}, {Bowman}, {Rogers},
  {Mozdzen}, {Cohen}, {Barkana}  \& {Mahesh}}{{Monsalve}
  et~al.}{2019}]{monsalve19}
{Monsalve} R.~A.,  {Fialkov} A.,  {Bowman} J.~D.,  {Rogers} A. E.~E.,
  {Mozdzen} T.~J.,  {Cohen} A.,  {Barkana} R.,   {Mahesh} N.,  2019, \mn@doi
  [\apj] {10.3847/1538-4357/ab07be}, \href
  {https://ui.adsabs.harvard.edu/abs/2019ApJ...875...67M} {875, 67}

\bibitem[\protect\citeauthoryear{{Mozdzen}, {Bowman}, {Monsalve}  \&
  {Rogers}}{{Mozdzen} et~al.}{2017}]{mozdzen17}
{Mozdzen} T.~J.,  {Bowman} J.~D.,  {Monsalve} R.~A.,   {Rogers} A.~E.~E.,
  2017, \mn@doi [\mnras] {10.1093/mnras/stw2696}, \href
  {https://ui.adsabs.harvard.edu/abs/2017MNRAS.464.4995M} {464, 4995}

\bibitem[\protect\citeauthoryear{{Mozdzen}, {Mahesh}, {Monsalve}, {Rogers}  \&
  {Bowman}}{{Mozdzen} et~al.}{2019}]{mozdzen19}
{Mozdzen} T.~J.,  {Mahesh} N.,  {Monsalve} R.~A.,  {Rogers} A.~E.~E.,
  {Bowman} J.~D.,  2019, \mn@doi [\mnras] {10.1093/mnras/sty3410}, \href
  {https://ui.adsabs.harvard.edu/abs/2019MNRAS.483.4411M} {483, 4411}

\bibitem[\protect\citeauthoryear{{Mu{\~n}oz} \& {Loeb}}{{Mu{\~n}oz} \&
  {Loeb}}{2018}]{munoz18}
{Mu{\~n}oz} J.~B.,  {Loeb} A.,  2018, \mn@doi [\nat]
  {10.1038/s41586-018-0151-x}, \href
  {https://ui.adsabs.harvard.edu/abs/2018Natur.557..684M} {557, 684}

\bibitem[\protect\citeauthoryear{{Parsons} et~al.,}{{Parsons}
  et~al.}{2010}]{parsons10}
{Parsons} A.~R.,  et~al., 2010, \mn@doi [\aj] {10.1088/0004-6256/139/4/1468},
  \href {https://ui.adsabs.harvard.edu/abs/2010AJ....139.1468P} {139, 1468}

\bibitem[\protect\citeauthoryear{{Patra}, {Subrahmanyan}, {Raghunathan}  \&
  {Udaya Shankar}}{{Patra} et~al.}{2013}]{patra13}
{Patra} N.,  {Subrahmanyan} R.,  {Raghunathan} A.,   {Udaya Shankar} N.,  2013,
  \mn@doi [Experimental Astronomy] {10.1007/s10686-013-9336-3}, \href
  {https://ui.adsabs.harvard.edu/abs/2013ExA....36..319P} {36, 319}

\bibitem[\protect\citeauthoryear{{Philip} et~al.,}{{Philip}
  et~al.}{2019}]{philip19}
{Philip} L.,  et~al., 2019, \mn@doi [Journal of Astronomical Instrumentation]
  {10.1142/S2251171719500041}, \href
  {https://ui.adsabs.harvard.edu/abs/2019JAI.....850004P} {8, 1950004}

\bibitem[\protect\citeauthoryear{Price et~al.,}{Price et~al.}{2018}]{price18}
Price D.~C.,  et~al., 2018, \mn@doi [Monthly Notices of the Royal Astronomical
  Society] {10.1093/mnras/sty1244}, 478, 4193

\bibitem[\protect\citeauthoryear{{Remazeilles}, {Dickinson}, {Banday},
  {Bigot-Sazy}  \& {Ghosh}}{{Remazeilles} et~al.}{2015}]{remazeilles15}
{Remazeilles} M.,  {Dickinson} C.,  {Banday} A.~J.,  {Bigot-Sazy} M.~A.,
  {Ghosh} T.,  2015, \mn@doi [\mnras] {10.1093/mnras/stv1274}, \href
  {https://ui.adsabs.harvard.edu/abs/2015MNRAS.451.4311R} {451, 4311}

\bibitem[\protect\citeauthoryear{{Schneider}, {Schmidt}  \& {Gunn}}{{Schneider}
  et~al.}{1991}]{schneider91}
{Schneider} D.~P.,  {Schmidt} M.,   {Gunn} J.~E.,  1991, \mn@doi [\aj]
  {10.1086/115914}, \href
  {https://ui.adsabs.harvard.edu/abs/1991AJ....102..837S} {102, 837}

\bibitem[\protect\citeauthoryear{{Shaver}, {Windhorst}, {Madau}  \& {de
  Bruyn}}{{Shaver} et~al.}{1999}]{shaver99}
{Shaver} P.~A.,  {Windhorst} R.~A.,  {Madau} P.,   {de Bruyn} A.~G.,  1999,
  \aap, \href {https://ui.adsabs.harvard.edu/abs/1999A&A...345..380S} {345,
  380}

\bibitem[\protect\citeauthoryear{{Shen}, {Anstey}, {de Lera Acedo}, {Fialkov}
  \& {Handley}}{{Shen} et~al.}{2021}]{shen21}
{Shen} E.,  {Anstey} D.,  {de Lera Acedo} E.,  {Fialkov} A.,   {Handley} W.,
  2021, \mn@doi [\mnras] {10.1093/mnras/stab429}, \href
  {https://ui.adsabs.harvard.edu/abs/2021MNRAS.503..344S} {503, 344}

\bibitem[\protect\citeauthoryear{{Sims} \& {Pober}}{{Sims} \&
  {Pober}}{2019}]{sims19}
{Sims} P.~H.,  {Pober} J.~C.,  2019, arXiv e-prints, \href
  {https://ui.adsabs.harvard.edu/abs/2019arXiv191003165S} {p. arXiv:1910.03165}

\bibitem[\protect\citeauthoryear{{Singh} \& {Subrahmanyan}}{{Singh} \&
  {Subrahmanyan}}{2019}]{singh19}
{Singh} S.,  {Subrahmanyan} R.,  2019, {The Redshifted 21 cm Signal in the
  EDGES Low-band Spectrum} (\mn@eprint {arXiv} {1903.04540}),
  \mn@doi{10.3847/1538-4357/ab2879}

\bibitem[\protect\citeauthoryear{{Singh} et~al.,}{{Singh}
  et~al.}{2018}]{singh18}
{Singh} S.,  et~al., 2018, \mn@doi [\apj] {10.3847/1538-4357/aabae1}, \href
  {https://ui.adsabs.harvard.edu/abs/2018ApJ...858...54S} {858, 54}

\bibitem[\protect\citeauthoryear{Skilling}{Skilling}{2006}]{skilling06}
Skilling J.,  2006, \mn@doi [Bayesian Anal.] {10.1214/06-BA127}, 1, 833

\bibitem[\protect\citeauthoryear{Slatyer \& Wu}{Slatyer \&
  Wu}{2018}]{slatyer18}
Slatyer T.~R.,  Wu C.-L.,  2018, \mn@doi [Physical Review D]
  {10.1103/physrevd.98.023013}, 98

\bibitem[\protect\citeauthoryear{{Sokolowski} et~al.,}{{Sokolowski}
  et~al.}{2015}]{sokolowski15}
{Sokolowski} M.,  et~al., 2015, \mn@doi [\pasa] {10.1017/pasa.2015.3}, \href
  {https://ui.adsabs.harvard.edu/abs/2015PASA...32....4S} {32, e004}

\bibitem[\protect\citeauthoryear{{Tauscher}, {Rapetti}  \& {Burns}}{{Tauscher}
  et~al.}{2020a}]{tauscher20a}
{Tauscher} K.,  {Rapetti} D.,   {Burns} J.~O.,  2020a, \mn@doi [\apj]
  {10.3847/1538-4357/ab9a3f}, \href
  {https://ui.adsabs.harvard.edu/abs/2020ApJ...897..132T} {897, 132}

\bibitem[\protect\citeauthoryear{Tauscher, Rapetti  \& Burns}{Tauscher
  et~al.}{2020b}]{tauscher20b}
Tauscher K.,  Rapetti D.,   Burns J.~O.,  2020b, \mn@doi [The Astrophysical
  Journal] {10.3847/1538-4357/ab9b2a}, 897, 175

\bibitem[\protect\citeauthoryear{{Trotta}}{{Trotta}}{2008}]{trotta08}
{Trotta} R.,  2008, \mn@doi [Contemporary Physics] {10.1080/00107510802066753},
  \href {https://ui.adsabs.harvard.edu/abs/2008ConPh..49...71T} {49, 71}

\bibitem[\protect\citeauthoryear{{Voytek}, {Natarajan}, {J{\'a}uregui
  Garc{\'\i}a}, {Peterson}  \& {L{\'o}pez-Cruz}}{{Voytek}
  et~al.}{2014}]{voytek14}
{Voytek} T.~C.,  {Natarajan} A.,  {J{\'a}uregui Garc{\'\i}a} J.~M.,  {Peterson}
  J.~B.,   {L{\'o}pez-Cruz} O.,  2014, {Probing the Dark Ages at z
  \raisebox{-0.5ex}\textasciitilde 20: The SCI-HI 21 cm All-sky Spectrum
  Experiment} (\mn@eprint {arXiv} {1311.0014}),
  \mn@doi{10.1088/2041-8205/782/1/L9}

\bibitem[\protect\citeauthoryear{{Wouthuysen}}{{Wouthuysen}}{1952}]{wouthuysen52}
{Wouthuysen} S.~A.,  1952, \mn@doi [\aj] {10.1086/106661}, \href
  {https://ui.adsabs.harvard.edu/abs/1952AJ.....57R..31W} {57, 31}

\bibitem[\protect\citeauthoryear{{Zaldarriaga}, {Furlanetto}  \&
  {Hernquist}}{{Zaldarriaga} et~al.}{2004}]{zaldarriaga04}
{Zaldarriaga} M.,  {Furlanetto} S.~R.,   {Hernquist} L.,  2004, \mn@doi [\apj]
  {10.1086/386327}, \href
  {https://ui.adsabs.harvard.edu/abs/2004ApJ...608..622Z} {608, 622}

\bibitem[\protect\citeauthoryear{{de Lera Acedo}}{{de Lera
  Acedo}}{2019}]{acedo19}
{de Lera Acedo} E.,  2019, in 2019 International Conference on Electromagnetics
  in Advanced Applications (ICEAA). pp 0626--0629

\bibitem[\protect\citeauthoryear{{de Oliveira-Costa}, {Tegmark}, {Gaensler},
  {Jonas}, {Landecker}  \& {Reich}}{{de Oliveira-Costa}
  et~al.}{2008}]{deoliveiracosta08}
{de Oliveira-Costa} A.,  {Tegmark} M.,  {Gaensler} B.~M.,  {Jonas} J.,
  {Landecker} T.~L.,   {Reich} P.,  2008, \mn@doi [\mnras]
  {10.1111/j.1365-2966.2008.13376.x}, \href
  {https://ui.adsabs.harvard.edu/abs/2008MNRAS.388..247D} {388, 247}

\bibitem[\protect\citeauthoryear{{van Haarlem} et~al.,}{{van Haarlem}
  et~al.}{2013}]{vanhaarlem13}
{van Haarlem} M.~P.,  et~al., 2013, \mn@doi [\aap]
  {10.1051/0004-6361/201220873}, \href
  {https://ui.adsabs.harvard.edu/abs/2013A&A...556A...2V} {556, A2}

\makeatother
\end{thebibliography}

%%%%%%%%%%%%%%%%%%%%%%%%%%%%%%%%%%%%%%%%%%%%%%%%%%

%%%%%%%%%%%%%%%%% APPENDICES %%%%%%%%%%%%%%%%%%%%%

%\appendix

%%%%%%%%%%%%%%%%%%%%%%%%%%%%%%%%%%%%%%%%%%%%%%%%%%

% Don't change these lines
\bsp	% typesetting comment
\label{lastpage}
\end{document}